\documentclass[twocolumn,secnumarabic,amssymb,nobibnotes,reprint,aps,prd]{revtex4-1}
\usepackage{graphicx}
\usepackage{color}
\usepackage{bm}

\begin{document}

\title{\boldmath Static Structures of the BCS-like Holographic Superfluid in AdS$_{4}$ Spacetime}

\author{Shanquan Lan}
\email{shanquanlan@mail.bnu.edu.cn}
\affiliation{Department of Physics, Beijing Normal University, Beijing, 100875, China}
\author{Wenbiao Liu}
\email{wbliu@bnu.edu.cn}
\affiliation{Department of Physics, Beijing Normal University, Beijing, 100875, China}
\author{Yu Tian}
\email{ytian@ucas.ac.cn}
\affiliation{School of Physics, University of Chinese Academy of Sciences, Beijing, 100049, China}
\affiliation{Shanghai Key Laboratory of High Temperature Superconductors, Shanghai, 200444, China}

\begin{abstract}
We investigate in detail the $m^{2}=0$ Abelian Higgs model in AdS$_{4}$, which is considered as the holographic dual of the most BCS-like superfluid. In homogeneous and isotropic superfluid solutions, we calculate the sound speeds, square of which approaches to $1/2$ with increasing chemical potential (lowering temperature). Then we present the single dark soliton solutions, which becomes thinner with increasing chemical potential. For the first time, we also find the interesting double and triple dark soliton solutions, which is unexpected and shows the possibility of more complicated static configurations. Finally, we investigate vortex solutions. For winding number $n=1$, the vortex becomes thinner with increasing chemical potential. At a given chemical potential, with increasing winding number, firstly the vortex becomes bigger and the charge density depletion becomes larger, then the charge density depletion settles down at a certain value and the growth of the vortex size is found to obey a scaling symmetry.
\end{abstract}

\maketitle
\flushbottom

\section{Introduction and Motivation}
\label{intro}

In AdS/CFT correspondence, an Abelian Higgs model coupled with gravity in $AdS_{4}$\cite{3h1} which exhibits spontaneous symmetry breaking of $U(1)$ symmetry has been extensively studied. Firstly, this model was considered to describe a holographic superconductor\cite{3h2}. In this case, the spatial components of the boundary $U(1)$ gauge fields can be turned on where droplet and vortex solutions have been discussed in Refs. \cite{montull2009holographic,albash2009vortex,maeda2010vortex,rozali2012holographic,haiqing2015}. Later on, interpretation of this model as a holographic superfluid can be found in Refs. \cite{herzog2009holographic,basu2009supercurrent,herzog2009sound,keranen2009dark,chesler2013holographic}. In this case, the spatial components of the boundary $U(1)$ gauge fields are turned off and yet consistent soliton and vortex solutions are found in Refs. \cite{Keranen2009ss,keranen2010inhomogeneous}. All in all, the model with different boundary conditions corresponds to superconductor or superfluid. In this paper, we will consider the holographic superfluid case.

In the model, the complex scalar field is massive with a coupling constant $m^{2}$. This parameter is related to the conformal dimension $\Delta$ of condensate by $\Delta_{\pm}=3/2\pm\sqrt{9/4+m^{2}}$\cite{klebanov1999ads} in four dimensional spacetime. From the equation, one can see $m^{2}>-9/4$ which is known as the Breitenlohner-Freedman bound\cite{breitenlohner1982positive}. What is more, if $m^{2}>0$, we have $\Delta_{-}<0$ and the scalar field goes to infinity at the boundary if the source is turned on which renders the model unstable. As a result, $m^{2}\leq0$ and case $m^{2}=0$ has a condensate with the maximum conformal dimension $\Delta_{+}=3$. {In Refs.\cite{Keranen2009ss,keranen2010inhomogeneous,keranen2011solitons}, after an investigation of this model for different $m^{2}$ from various aspects, the authors suggest that the increasing of conformal dimension corresponds to the crossover from BEC superfluid to BCS superfluid. Thus our model corresponds to the BCS superfluid with the most loosely bounded cooper pairs of fermions. As we know the BCS limit is currently not accessible to ultra-cold gas experiments\cite{Pitaevskii2013snake}, so a theoretical investigation of this ``extreme'' BCS-like holographic superfluid is valuable.}

Many characteristics of superfluid are interesting, such as the turbulence features that have been investigated in Refs.\cite{chesler2013holographic,Ewerz2014tua,du2014holographic,lan2016towards} for different kinds of holographic superfluid. But before an investigation of this extreme BCS-like holographic superfluid's dynamic features, first we shall know its static structures which is also interesting and important by itself. Actually, some of the static features are already known. One can find the phase transition diagram in Ref.\cite{horowitz2008holographic} and the single dark soliton, vortex solutions in Ref.\cite{keranen2011solitons}. But a detailed and deeper investigation are still necessary. In this paper, besides the three kinds of static structures mentioned above (the homogeneous superfluid, single dark soliton and vortex) constructed in the infalling Eddington Coordinates, we also find some more complicated structures and investigate their features.

For the homogeneous and isotropic superfluid solutions, we present the sound velocity of the superfluid at different chemical potential. The square of sound speed approaches to $1/2$ with increasing chemical potential. This is consistent with the argument in Ref.\cite{yarom2009fourth} where for the conformal fluids at zero temperature the square of sound speed is $1/2$.

For the single soliton solutions, we study the features of their charge density and healing length varying with respect to chemical potential. Then, for the first time, we find the multiple (double and triple) dark soliton solutions. Somewhat unexpectedly, those multiple dark solitons seem to balance at quite arbitrary positions, at least to very high precision numerically. Nevertheless, the nature of such multiple soliton solutions at finite temperature and their possible applications worth further investigation.

For the winding number $n=1$ vortex, we study the features of their charge density and healing length varying with respect to chemical potential. Then we study the vortices with different winding numbers at a certain chemical potential. Interestingly, we find that there is a scaling symmetry when the winding number is very large, which governs the growth of the vortex size with respect to the winding number.

The paper is organized as follows. In the next section, we develop the $m^{2}=0$ holographic superfluid model in the infalling Eddington Coordinates. Then we present the homogeneous and isotropic superfluid solutions, the dark soliton solutions and the vortex solutions in the next three sections. In the end, we summarize what we have found with some discussions. By the way, we have also find the multiple soliton solutions for the $m^{2}=-2$ holographic superfluid model and put it in the Appendix.

\section{Holographic Setup}
\label{setup}

As argued in Refs.\cite{3h1,3h2}, a holographic dual of a two dimensional superfluid is provided by Abelian Higgs model in asymptotically AdS$_{4}$ spacetime. The action is
\begin{equation}
S=\frac{1}{16\pi{G}}\int_{\mathcal{M}}\mathnormal{d}^{4}x\sqrt{-g}(R+\frac{6}{L^{2}}+\frac{1}{q^{2}}\mathcal{L}_{matter}).
\end{equation}
Here $G$ is the Newton's constant, and the matter Lagrangian reads
\begin{equation}
\mathcal{L}_{matter}=-\frac{1}{4}F_{ab}F^{ab}-|D\Psi|^{2}-m^{2}|\Psi|^{2},
\end{equation}
where $D=\nabla-iA$ with $\nabla$ the covariant derivative compatible to the metric.

In this paper, we will work in the probe limit(the backreaction of the matter fields onto gravity are ignored).
According to the holographic dictionary, the temperature of a black hole in the above model corresponds to the temperature of the superfluid. Since we are considering the superfluid at finite temperature, so we take the Schwarzschild black brane solution as our background geometry. In the infalling Eddington Coordinates, the metric is written as
\begin{eqnarray}
ds^{2}=\frac{L^{2}}{z^{2}}(-f(z)dt^{2}-2dt dz+dx^{2}+dy^{2}),
\end{eqnarray}
where the factor $f(z)=1-(\frac{z}{z_{h}})^{3}$ with $z=z_{h}$ the horizon and $z=0$ the AdS boundary. As alluded to above, the dual boundary system is placed at Hawking temperature, which is given by
\begin{equation}
T=\frac{3}{4\pi{z_{h}}}.
\end{equation}
On top of this background geometry, the equations of motion for the matter fields can then be written as
\begin{equation}
D_{a}D^{a}\Psi-m^{2}\Psi=0,\nabla_{a}F^{ab}=i(\overline{\Psi}D^{b}\Psi-\Psi\overline{D^{b}\Psi}),
\end{equation}
here the overbar denotes complex conjugate.
In what follows, we will take the units in which $L=1,16\pi Gq^{2}=1$, and $z_{h}=1$. Working with the case of $m^{2}=0$, the asymptotic solution of $A$ and $\Psi$ near the AdS boundary can be expanded as
\begin{equation}
A_{\mu}=a_{\mu}+b_{\mu}z+o(z),\Psi=\psi_{0}+\psi{z^{3}}+o(z).
\end{equation}
Again, according to the holographic dictionary, the expectation values of the boundary quantum field theory operators $j^{\mu}$ (currents) and $O$ (condensate) can be obtained explicitly as the variation of renormalized bulk on-shell action with respect to the sources $a_{\mu}$ and $\psi_{0}$ respectively
\begin{eqnarray}
\langle j^{\mu}\rangle&=&\frac{\delta{S_{onshell}}}{\delta{a_{\mu}}}=\lim_{z\rightarrow 0}\sqrt{-g}F^{z\mu},\nonumber\\
\langle O\rangle&=&\frac{\delta{S_{onshell}}}{\delta{\psi_{0}}}=-3\overline{\psi}.
\end{eqnarray}

Choosing the axial gauge $A_{z}=0$ to fix the $U(1)$ gauge fields, the above equations of motion are rewritten as
\begin{eqnarray}
&\,&2(1-z\partial_{z})\partial_{t}\Psi=(2+z^{3})\partial_{z}\Psi+2iA_{t}\Psi-iz\partial_{z}A_{t}\Psi\nonumber\\
&\,&-2izA_{t}\partial_{z}\Psi-z(\overrightarrow{\partial}-i\overrightarrow{A})\cdot(\overrightarrow{\partial}-i\overrightarrow{A})\Psi-zf\partial_{z}^{2}\Psi,
\end{eqnarray}
\begin{eqnarray}\label{constraintequation}
0=z^{2}(-\partial_{z}^{2}A_{t}+\partial_{z}\overrightarrow{\partial}\cdot\overrightarrow{A})+i(\overline{\Psi}\partial_{z}\Psi-\Psi\partial_{z}\overline{\Psi}),
\end{eqnarray}
\begin{eqnarray}
&\,&2z^{2}\partial_{z}\partial_{t}A_{x}=z^{2}(\partial_{x}\partial_{z}A_{t}+\partial_{y}^{2}A_{x}-\partial_{x}\partial_{y}A_{y}-3z^{2}\partial_{z}A_{x}\nonumber\\
&\,&+f\partial_{z}^{2}A_{x})-(i\overline{\Psi}\partial_{x}\Psi-i\Psi\partial_{x}\overline{\Psi}+2A_{x}\Psi^{2}),
\end{eqnarray}
\begin{eqnarray}
&\,&2z^{2}\partial_{z}\partial_{t}A_{y}=z^{2}(\partial_{y}\partial_{z}A_{t}+\partial_{x}^{2}A_{y}-\partial_{x}\partial_{y}A_{x}-3z^{2}\partial_{z}A_{y}\nonumber\\
&\,&+f\partial_{z}^{2}A_{y})-(i\overline{\Psi}\partial_{y}\Psi-i\Psi\partial_{y}\overline{\Psi}+2A_{y}\Psi^{2}),
\end{eqnarray}
\begin{eqnarray}
&\,&z^{2}(\partial_{x}\partial_{t}A_{x}+\partial_{y}\partial_{t}A_{y}+\partial_{z}\partial_{t}A_{t})+i(\overline{\Psi}\partial_{t}\Psi-\Psi\partial_{t}\overline{\Psi})\nonumber\\
&\,&=z^{2}(\partial_{x}^{2}A_{t}+\partial_{y}^{2}A_{t}+f\partial_{z}\overrightarrow{\partial}\cdot\overrightarrow{A})-2A_{t}\Psi^{2}\nonumber\\
&\,&+if(\overline{\Psi}\partial_{z}\Psi-\Psi\partial_{z}\overline{\Psi}).
\end{eqnarray}
The first equation of motion will contribute two independent equations for the scalar field is complex. As a result, there are six equations above. But we have chosen $A_{z}=0$ gauge, so only five equations are independent.

As a note, we consider the situation with following boundary conditions,
\begin{eqnarray}
\psi_{0}=0,\qquad a_{t}=\mu.
\end{eqnarray}
 Here $\mu$ is chemical potential, source of particle density $-b_{0}=\rho$. When fixing the black hole temperature $T$ by rescaling, the only scale parameter is $\mu$, which is inversely proportional to $T$. To solve the above equations, we use the Chebyshev and the Fourier pseudo-spectral methods\cite{guo2016modave}.

\section{homogeneous superfluid and sound speed}
\label{homosolution}

In this section, we reproduce the homogeneous and isotropic superfluid solutions\cite{horowitz2008holographic} in the infalling Eddington Coordinates. Then we present the sound velocity of the superfluid at different chemical potential. Same as predicted in Ref.\cite{yarom2009fourth}, the square of sound speed approaches to 1/2 with increasing the chemical potential in our numerical results.

\begin{figure}
\begin{center}
\includegraphics[scale=0.6]{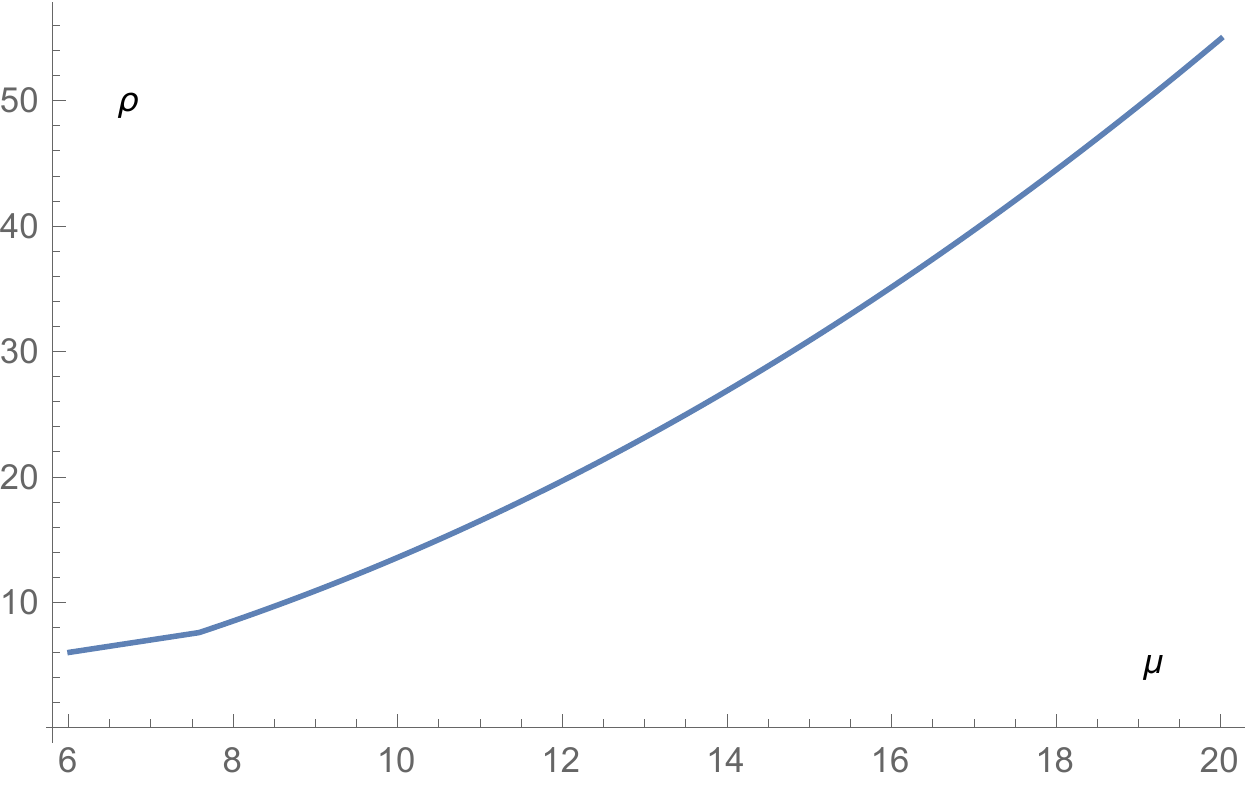}
\includegraphics[scale=0.6]{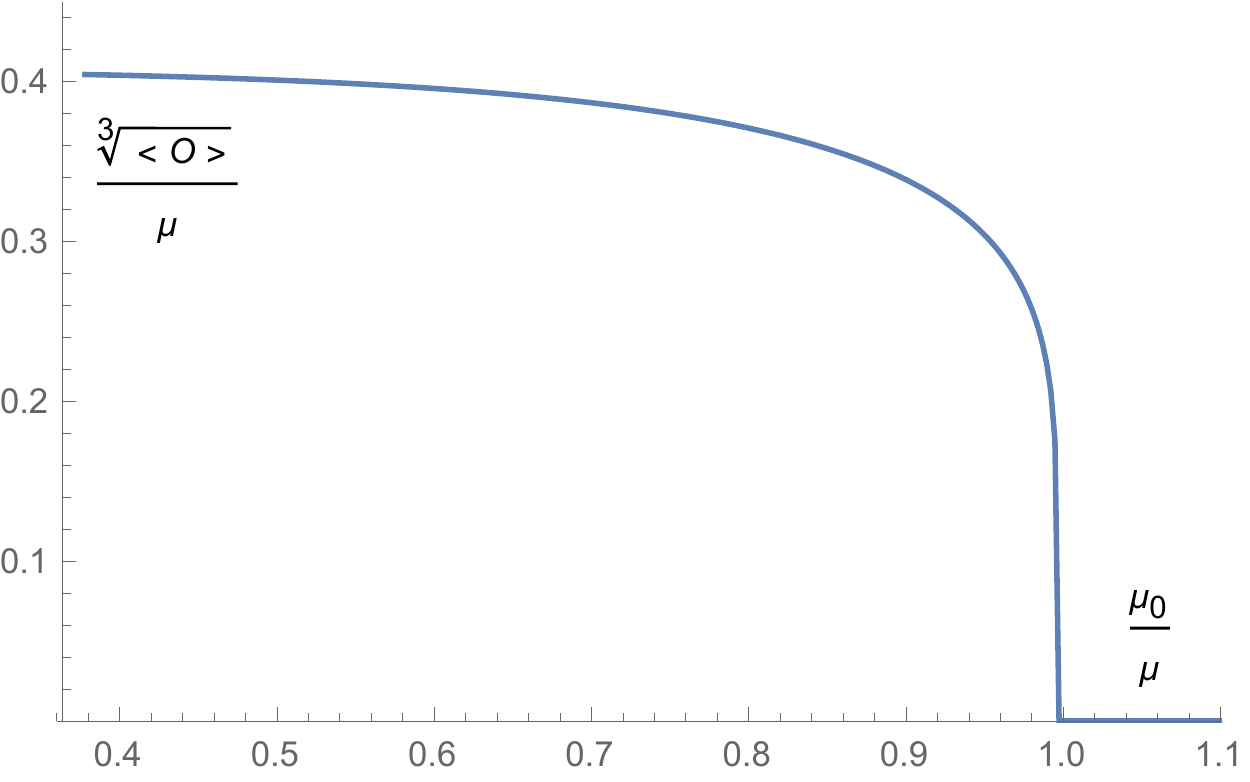}
\end{center}
\caption{The up panel shows the relation between chemical potential $\mu$ and charge density $\rho$, while the bottom panel shows the corresponding condensation with respect to temperature (inverse of the chemical potential $\mu$). The critical chemical potential is $\mu_{c}=7.56$ and the critical charge density is $\rho_{c}=7.56$.}\label{fig1}
\end{figure}

Considering the homogeneous, isotropic and static conditions, the matter fields only involve of $\Psi(z)$ and $A_{t}(z)$. Rewrite the complex scalar field $\Psi(z)=\phi(z)e^{i\varphi(z)}$ (which is also done in the following sections), the independent three equations of motion become
\begin{eqnarray}
zf\partial_{z}^{2}\phi-(2+z^{3})\partial_{z}\phi-2zA_{t}\phi\partial_{z}\varphi-zf\phi(\partial_{z}\varphi)^{2}=0,
\end{eqnarray}
\begin{eqnarray}
A_{t}+f\partial_{z}\varphi=0,
\end{eqnarray}
\begin{eqnarray}
z^{2}\partial_{z}^{2}A_{t}+2\phi^{2}\partial_{z}\varphi=0.
\end{eqnarray}
From the second equation, we find $A_{t}(z=1)=0$ and without loss of generality we set $\varphi(z=1)=0$. Together with the boundary conditions in Section \ref{setup}, the above equations of motion can be solved for different $\mu$. The numerical results show that, when $\mu$ is small, the solution is trivial $\phi=0$, while passing through the critical point $\mu_{c}=7.56$, $\phi$ begins to condense. This process denotes that when increasing $\mu$ (lowering the temperature $T$), normal fluid condense to superfluid through a phase transition. This picture can be clearly seen in the phase transition graph Fig.\ref{fig1}.

\begin{figure}
\begin{center}
\includegraphics[scale=0.6]{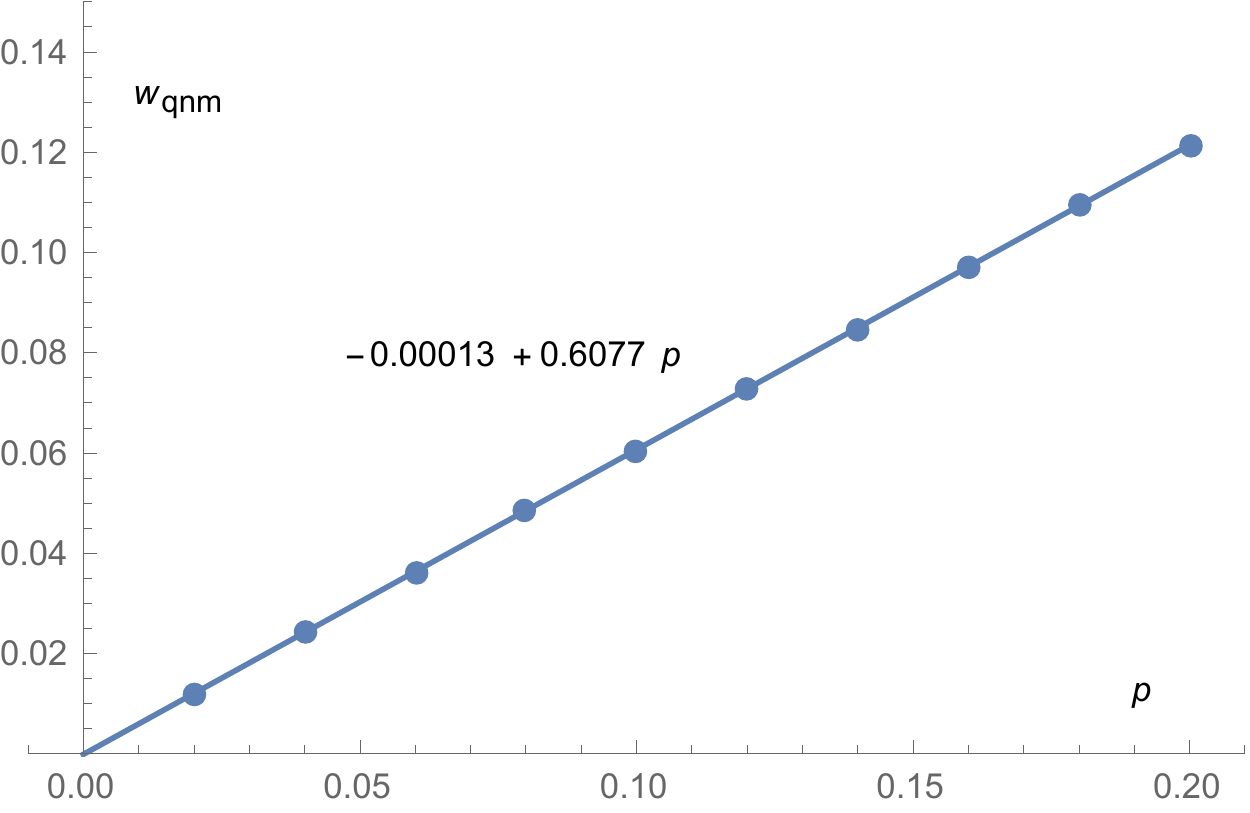}
\end{center}
\caption{The dispersion relation for the Goldstone mode in superfluid phase at $\mu=10$.}\label{fig2}
\end{figure}

\begin{figure}
\begin{center}
\includegraphics[scale=0.6]{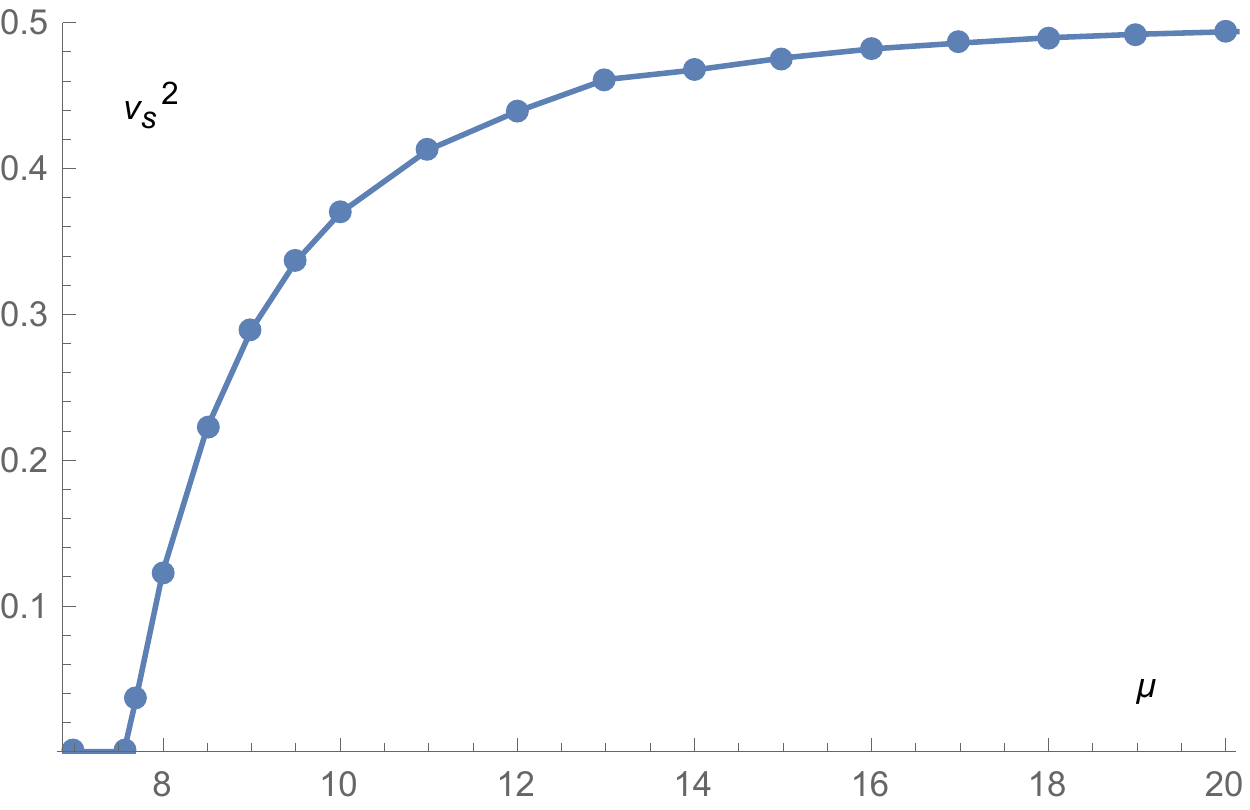}
\end{center}
\caption{The sound speed square $\upsilon_{s}^{2}$ varies with respect to chemical potential $\mu$. $\upsilon_{s}^{2}$ approaches $1/2$ with increasing $\mu$ (lowering temperature).}\label{fig3}
\end{figure}

With the above background solutions of $\Psi(z)=\phi(z){e}^{i\varphi(z)}=\Psi_{r}(z)+i\,\Psi_{i}(z)$ and $A_{t}(z)$, we are ready to calculate the sound speed of the superfluid system by the linear response theory\cite{amado2009hydrodynamics,du2015dynamical}. To do this, we assume the perturbation bulk fields take the following forms
\begin{eqnarray}
&\,&\delta\Psi_{r}=\delta\Psi_{r}(z)e^{-iwt+ipx},\delta\Psi_{i}=\delta\Psi_{i}(z)e^{-iwt+ipx},\nonumber\\
&\,&\delta A_{t}=\delta A_{t}(z)e^{-iwt+ipx},\delta A_{x}=\delta A_{x}(z)e^{-iwt+ipx}.
\end{eqnarray}
Then four independent perturbation equations (the extra fifth Eq.(\ref{constraintequation}) is not considered here) can be written as
\begin{eqnarray}
&\,&0=(2iw+p^{2}z+(2+z^{3}-2iwz)\partial_{z}-zf\partial_{z}^{2})\delta\Psi_{r}\nonumber\\
&\,&+(z\partial_{z}A_{t}-2A_{t}+2zA_{t}\partial_{z})\delta\Psi_{i}\nonumber\\
&\,&+(2z\partial_{z}\Psi_{i}-2\Psi_{i}+z\Psi_{i}\partial_{z})\delta A_{t}\nonumber\\
&\,&-ipz\Psi_{i}\delta A_{x},
\end{eqnarray}
\begin{eqnarray}
&\,&0=(2A_{t}-z\partial_{z}A_{t}-2zA_{t}\partial_{z})\delta\Psi_{r}\nonumber\\
&\,&+(2iw+p^{2}z+(2+z^{3}-2iwz)\partial_{z}-zf\partial_{z}^{2})\delta\Psi_{i}\nonumber\\
&\,&+(2\Psi_{r}-2z\partial_{z}\Psi_{r}-z\Psi_{r}\partial_{z})\delta A_{t}\nonumber\\
&\,&+ipz\Psi_{r}\delta A_{x},
\end{eqnarray}
\begin{eqnarray}
&\,&0=(2iw\Psi_{i}+2f\Psi_{i}\partial_{z}-2f\partial_{z}\Psi_{i}-4A_{t}\Psi_{r})\delta\Psi_{r}\nonumber\\
&\,&+(2f\partial_{z}\Psi_{r}-2iw\Psi_{r}-2f\Psi_{r}\partial_{z}-4A_{t}\Psi_{i})\delta\Psi_{i}\nonumber\\
&\,&+(iwz^{2}\partial_{z}-p^{2}z^{2}A_{t}-2(\Psi_{r}^{2}+\Psi_{i}^{2}))\delta A_{t}\nonumber\\
&\,&+(ipz^{2}f\partial_{z}-wpz^{2})\delta A_{x},
\end{eqnarray}
\begin{eqnarray}
&\,&0=-2ip\Psi_{i}\delta\Psi_{r}+2ip\Psi_{r}\delta\Psi_{i}+ipz^{2}\partial_{z}\delta A_{t}\nonumber\\
&\,&+((2iwz^{2}-3z^{4})\partial_{z}+z^{2}f\partial_{z}^{2}-2(\Psi_{r}^{2}+\Psi_{i}^{2}))\delta A_{x}.
\end{eqnarray}
The above linear perturbation equations are required to have Dirichlet boundary conditions at the AdS boundary and regular conditions at horizon. Then cast them into the form $\mathcal{L}(w,p)v=0$ with $v$ the perturbation fields evaluated at the grid points by pseudo-spectral method. The quasi-normal modes $w_{qnm}$ are obtained by the condition $\det[\mathcal{L}(w,p)]=0$ ($w$ is complex and its real part is the quasi-normal mode). Thus the dispersion relation can be obtained in Fig.\ref{fig2} and the sound speed $\upsilon_{s}$ can be obtained by the fitting formula $w_{qnm}=\upsilon_{s}p$. In Fig.\ref{fig3}, we show the square of the sound speed which approaches $1/2$ with respect to chemical potential. This maximum sound speed $1/2$ feature is shared by conformal fluids at zero temperature\cite{herzog2009sound,yarom2009fourth}.

\section{inhomogeneous structures: dark soliton}
\label{solition}

Following the work in Refs.\cite{keranen2009dark,Keranen2009ss} where dark soliton solutions are constructed in the $m^{2}=-2$ Abelian Higgs holographic model, here we reproduce the single dark soliton solutions for $m^{2}=0$ case\cite{keranen2011solitons} in the infalling Eddington Coordinates. Then we study the charge density and healing length of single dark soliton solutions, and present multiple (double and triple) dark soliton solutions.

Single dark soliton solutions and multiple dark soliton solutions with several single dark soliton paralleling to each other have one dimensional translation invariance symmetry. Thus we can assume the fields are independent of $y$ coordinate, and $A_{y}=0$. As we are looking for static solutions, all the fields are independent of $t$ coordinate and the currents in the gravity system should be zero. As a result, rewrite the complex scalar field $\Psi(x,z)=\phi(x,z)e^{i\varphi(x,z)}$, the currents in Eq.(5) will contribute two constraint equations
\begin{equation}
A_{x}(x,z)-\partial_{x}\varphi(x,z)=0,
\end{equation}
\begin{equation}
A_{t}(x,z)+f(z)\partial_{z}\varphi(x,z)=0.
\end{equation}
By using these two constraint equations, the equations of motion in Section \ref{setup} will be dramatically simplified as
\begin{eqnarray}
zf(\partial_{z}\varphi)^{2}\phi-(2+z^{3})\partial_{z}\phi+zf\partial_{z}^{2}\phi+z\partial_{x}^{2}\phi=0,
\end{eqnarray}
\begin{eqnarray}
&\,&2\partial_{z}\varphi\phi^{2}+z^{2}(6z\partial_{z}\varphi+6z^{2}\partial_{z}^{2}\varphi\nonumber\\
&\,&-f\partial_{z}^{3}\varphi-\partial_{z}\partial_{x}^{2}\varphi)=0.
\end{eqnarray}

Ideally, the size of $x$ direction should be infinite. But in numerical method, a cut off is needed. We will consider the system in a box of size $1\times 2L$ in the $z$ and $x$ directions respectively. The boundary conditions are introduced as follows. At $z=0$, $\phi(z=0)=0$ (the condensate source is turned off), $\partial_{z}\varphi(z=0)=-\mu$ (setting the chemical potential which is introduced as Eq.(23)). At $z=1$, we put regular boundary condition for $\phi$, and $\varphi(z=1)=0$ (which is introduced as Eq.(23) by the same argument as that in the above section). At $x=-L$ and $x=L$, we will introduce Neumann boundary conditions for both $\phi$ and $\varphi$ fields. With the equations of motion and boundary conditions set up, we can use relaxation methods ( Newton iteration method, or Gauss-Seidel iteration method) to find solutions. First, we choose seed configurations of $\phi(x,z)$ and $\varphi(x,z)$. Then relax these configurations towards the solutions.

\subsection{single dark soliton solutions}
\label{singlesolition}

In this subsection, we study in detail the single dark soliton solutions for different chemical potential. The seed configurations are chosen as $\phi(x,z)=a\tanh(x)$ and $\varphi(x,z)=b$, where $a,b$ are constants. Then relax these configurations towards the solutions.

\begin{figure}
\begin{center}
\includegraphics[scale=0.76]{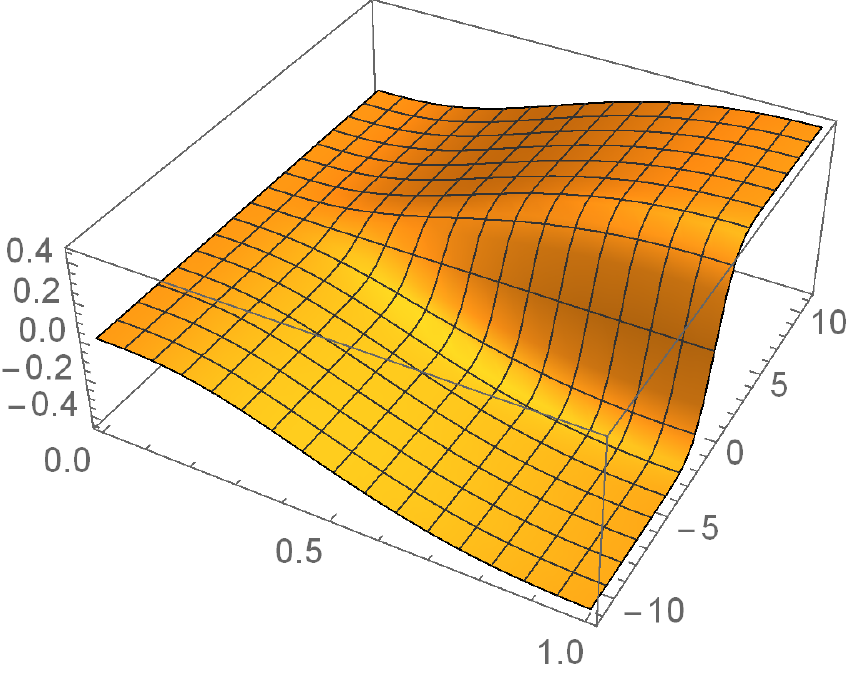}
\includegraphics[scale=0.8]{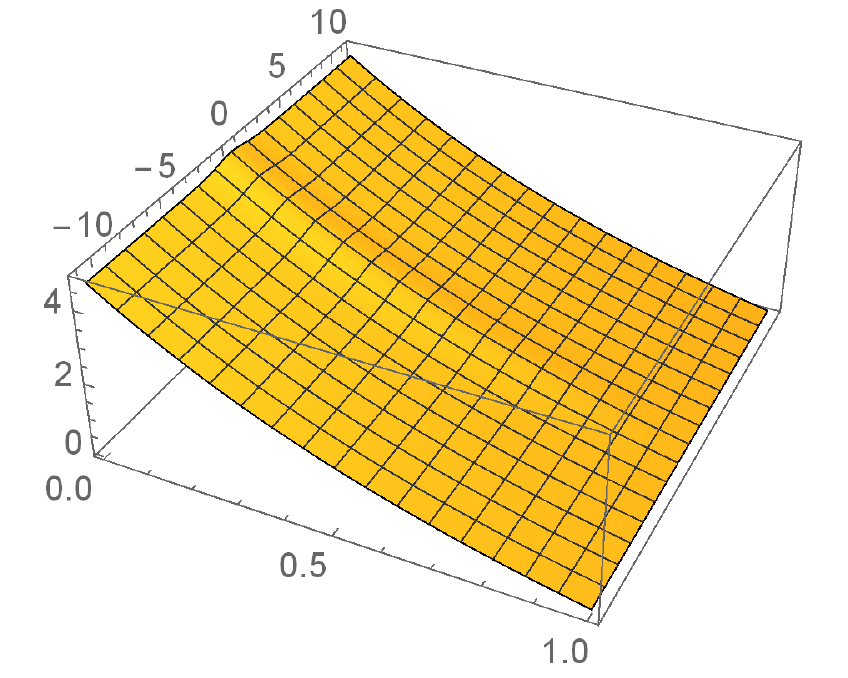}
\includegraphics[scale=0.8]{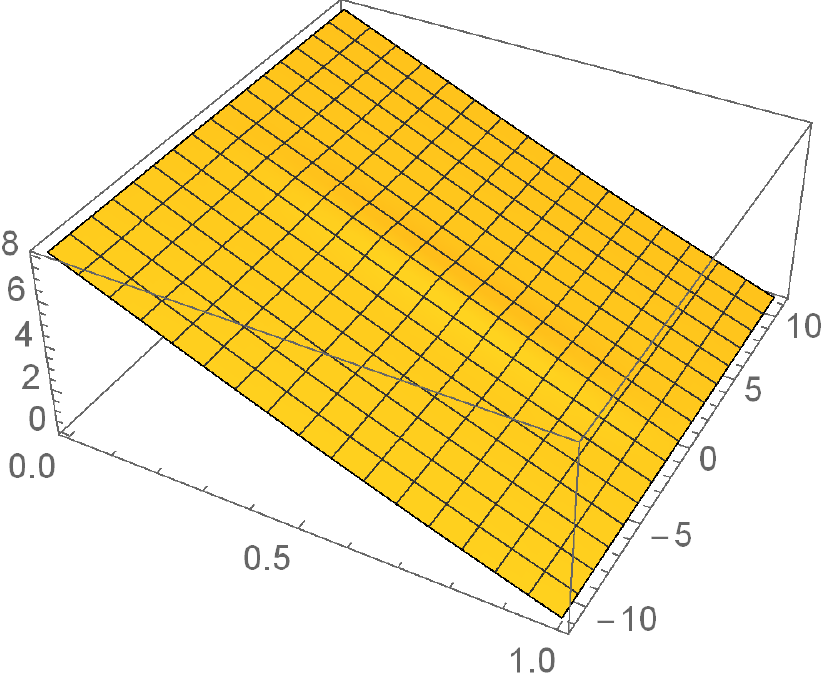}
\includegraphics[scale=0.8]{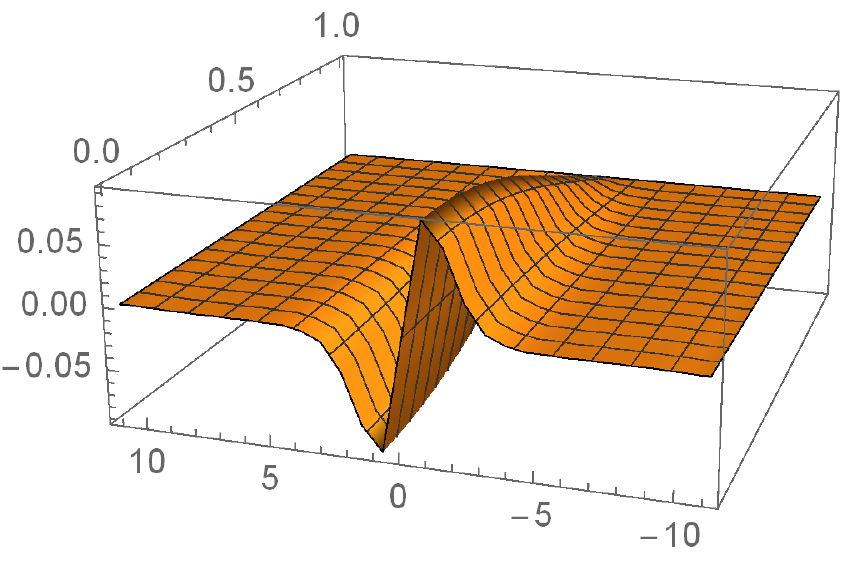}
\end{center}
\caption{The bulk configurations of the fields $\phi(x,z),\varphi(x,z),A_{t}(x,z),A_{x}(x,z)$ from up to down respectively for $\mu=8$.}\label{fig4}
\end{figure}

\begin{figure}
\begin{center}
\includegraphics[scale=0.5]{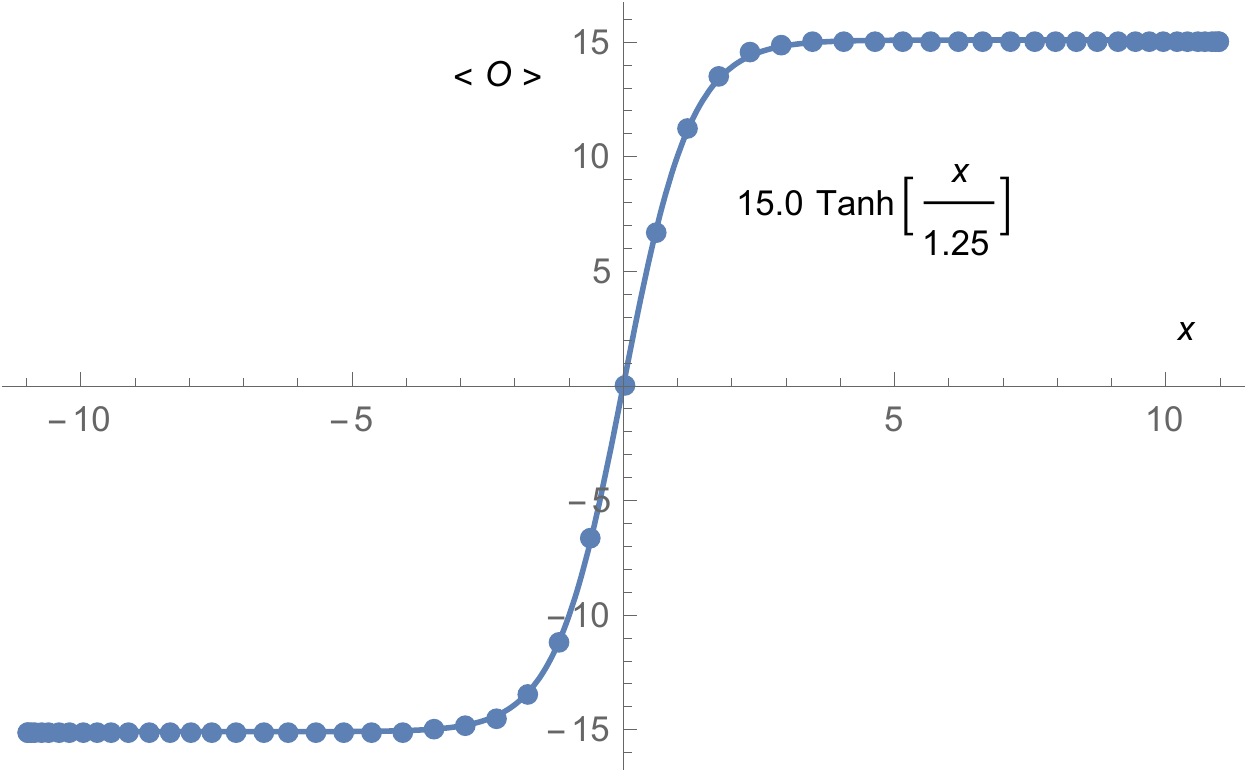}
\includegraphics[scale=0.5]{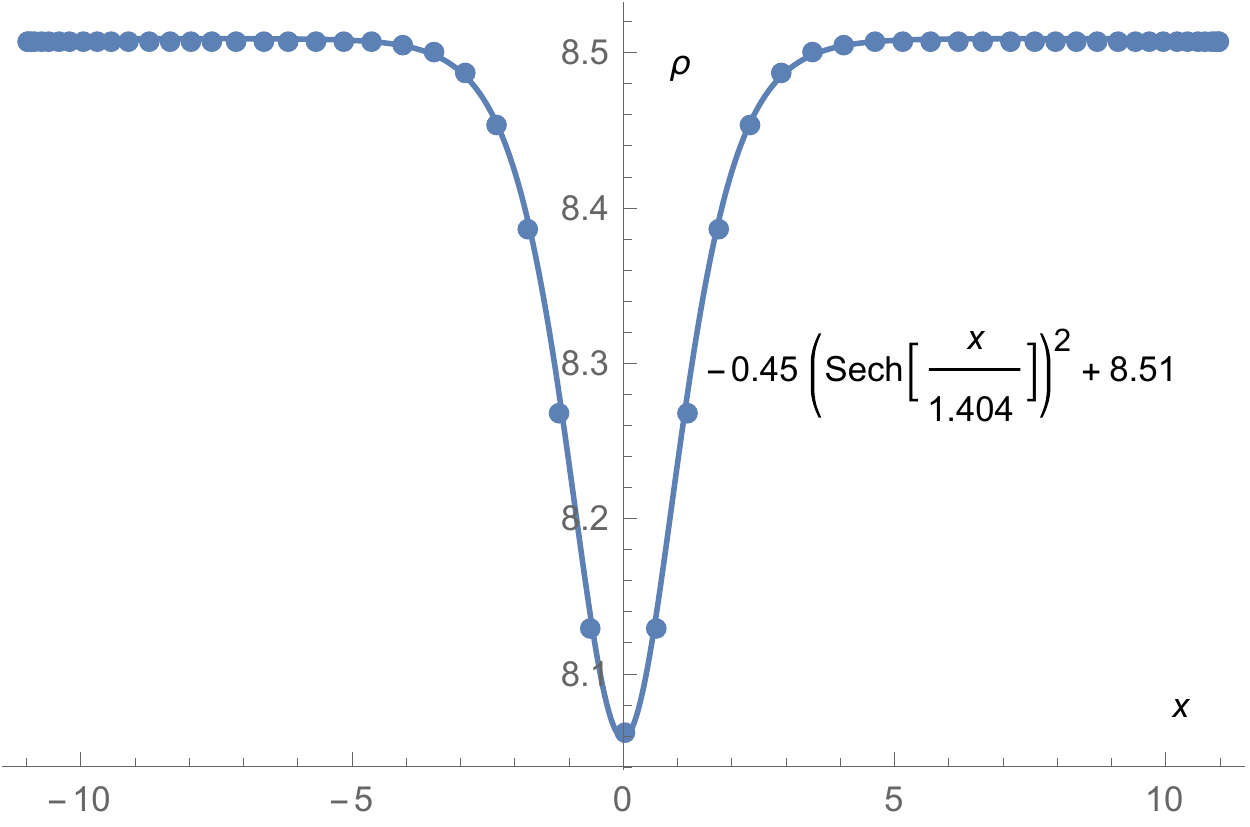}
\end{center}
\caption{Condensate and charge density of soliton as functions of $x$ at $\mu=8$. The solid lines are fitting graphes of respective functions.}\label{fig5}
\end{figure}

For chemical potential $\mu=8$, the numerical results of the bulk fields configurations are shown in Fig.\ref{fig4}.
Condensate and charge density as functions of $x$ are shown in Fig.\ref{fig5}.
The condensate is well fitted by function
\begin{equation}
\langle O \rangle (x)=15.0\tanh(x/1.25),
\end{equation}
where $15.0$ is condensate for homogeneous and isotropic solution with the same chemical potential and 1.25 should be thought of as healing length of the soliton. The charge density is well fitted by function
\begin{equation}
\rho=-0.45\,sech^{2}(x/1.404)+8.51,
\end{equation}
where $8.51$ is charge density for homogeneous and isotropic solution with the same chemical potential, $1.404$ should also be thought of as healing length of the soliton and $0.45$ is the charge density depletion. Thus there are two healing lengths for a holographic soliton. Compare the results with the soliton solutions of the Gross-Pitaevskii equation (GPE)\cite{Keranen2009ss} in condensed matter physics, it is also found that the forms of the fitting functions are the same, but the charge density in the soliton core is zero and there is only one soliton healing length for the GPE case.

For different chemical potential, the parameters which characterize soliton are listed in Table.\ref{table1}. As we increase the chemical potential (lower the temperature), (1)the condensate increases dramatically, (2)the depletion of charge density increases and the two healing lengths (width of the soliton) decreases, and as a result, the solitons become thinner and smaller but deeper, (3)$\delta\rho/\rho_{max}$ increases fast near the critical point but slows down later, for all $\mu$, this ratio is small which means the depletion of soliton is shallow in charge density and it is not likely to increase to 1 (the charge density at the soliton core is not likely to be 0).
\begin{table}
\begin{center}
	\begin{tabular}{|c|c|c|c|c|c|c|c|}
		\hline
		\,\,\,\,$\mu$\,\,\,\, & \,\,$\langle O \rangle_{max}$ \,\,& \,\,\,$\rho_{max}$\,\,\, & \,\,\,\,$\delta\rho$ \,\,\,\,&\,\,$\delta\rho/\rho_{max}$ \,\,& \,\,\,$\epsilon_{c}$ \,\,\,& \,\,\,\,$\epsilon_{\rho}$\,\,\,\, \\
	 \hline
  8 & 15.0 & 8.51 & 0.45 &0.053 & 1.250 & 1.404 \\
  \hline
  9 & 34.2 & 10.9 & 1.43 &0.131 & 0.630 & 0.806 \\
  \hline
  10 & 54.5 & 13.6 & 2.27 &0.167 & 0.456 & 0.638\\
  \hline
  11 & 77.9 & 16.5 & 3.17 &0.192 & 0.367 & 0.541 \\
  \hline
  12 & 105.4 & 19.7 & 4.07 &0.207 & 0.313 & 0.475 \\
  \hline
  13 & 137.5 & 23.1 & 5.06 &0.219 & 0.272 & 0.428 \\
  \hline
  14 & 174.7 & 26.9 & 6.06 & 0.225 &0.244 & 0.388 \\
  \hline
  15 & 217.3 & 30.9 & 7.07 & 0.229 &0.227 & 0.361 \\
  \hline
  16 & 266.0 & 35.2 & 8.16 & 0.232 &0.211 & 0.333 \\
  \hline
  17 & 321.2 & 39.7 & 9.35 & 0.236 &0.194 & 0.310 \\
  \hline
  18 & 383.2 & 44.5 & 10.64 & 0.239 &0.179 & 0.291 \\
  \hline
  19 & 452.6 & 49.6 & 12.0 & 0.242 &0.165 & 0.274 \\
  \hline
  20 & 530.0 & 55.0 & 13.5 & 0.246 &0.152 & 0.260 \\
  \hline
	\end{tabular}
\end{center}
\caption{The relation between chemical potential $\mu$ and max condensate $\langle O \rangle_{max}$, max charge density $\rho_{max}$, charge density depletion $\delta \rho$, healing length $\epsilon_{c}$ for condensate, healing length $\epsilon_{\rho}$ for charge density.}\label{table1}
\end{table}

\subsection{multiple dark soliton solutions}
\label{manysolition}

It is known that the Gross-Pitaevskii equation also has multiple dark soliton solutions, but such solutions cannot be static in general, due to the force of interaction between these solitons (see, e.g. Ref.\cite{Frantzeskakis}). Since the dark soliton is a localized structure, static configurations of multiple dark solitons with large enough distance between them act as important approximate solutions of the theory, which are in principle unstable but can have arbitrarily long lifetime. Therefore, it is interesting to explore similar solutions at finite temperature by holography, approximate or exact. However, as far as we know, there is no such exploration up to now.

In this subsection, we initiate such exploration by finding the double and triple parallel dark soliton solutions in the holographic superfluids. Interestingly, it seems that these solutions are exact, at least in the conditions that we consider. Actually, these solutions can be found both in the $m^2=-2$ case and the $m^2=0$ case (and probably in other cases as well), but we will focus on the $m^2=0$ case here and leave the $m^2=-2$ case in the Appendix.

\begin{figure}
\begin{center}
\includegraphics[scale=0.7]{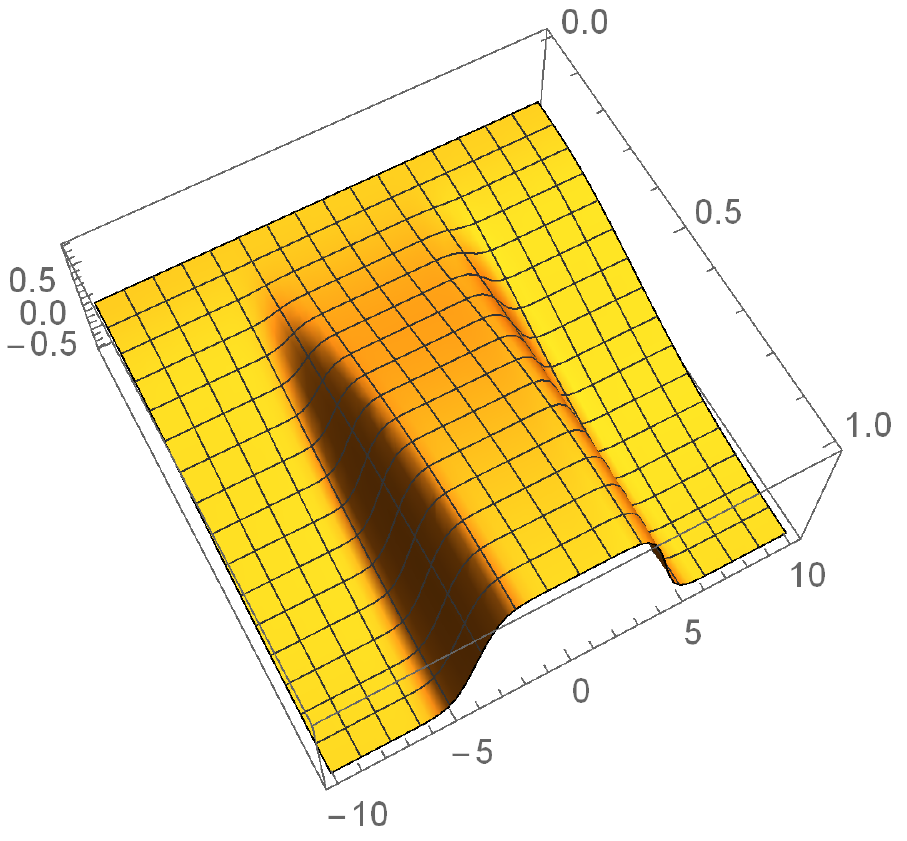}
\includegraphics[scale=0.7]{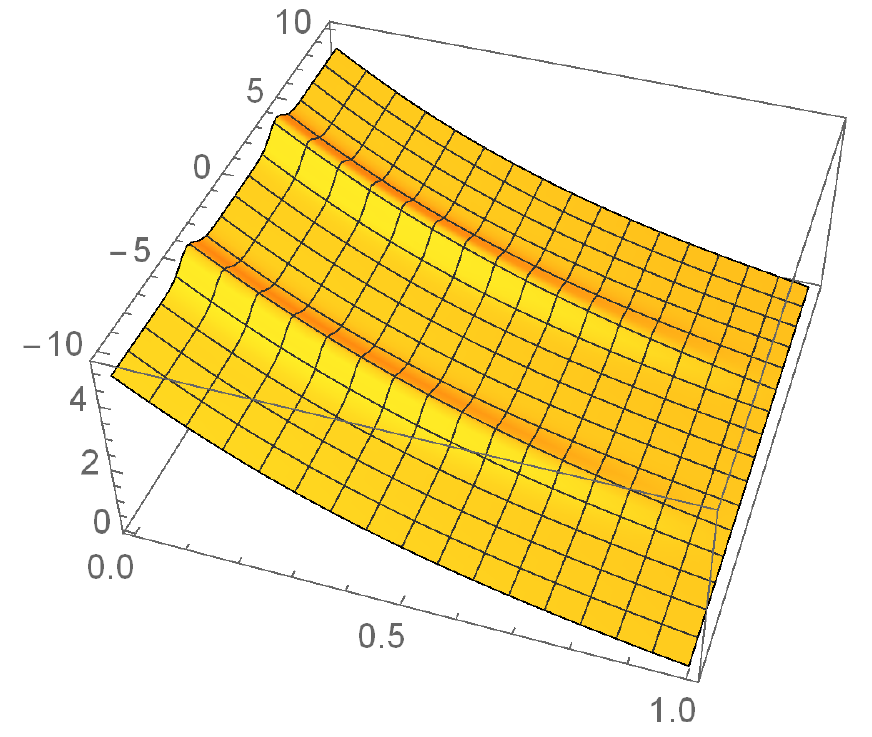}
\end{center}
\caption{The bulk configurations of the double soliton fields $\phi(x,z),\varphi(x,z)$ from up to down respectively for $\mu=9$.}\label{fig6}
\end{figure}

\begin{figure}
\begin{center}
\includegraphics[scale=0.5]{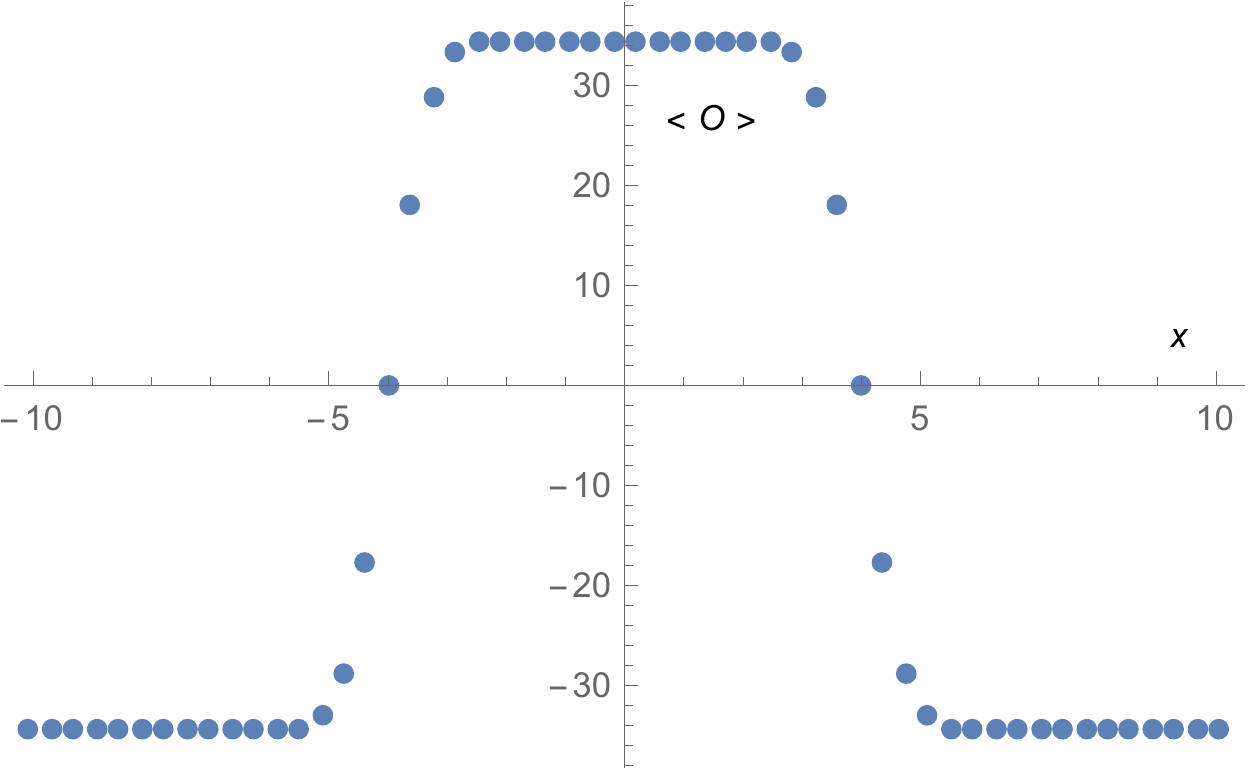}
\includegraphics[scale=0.5]{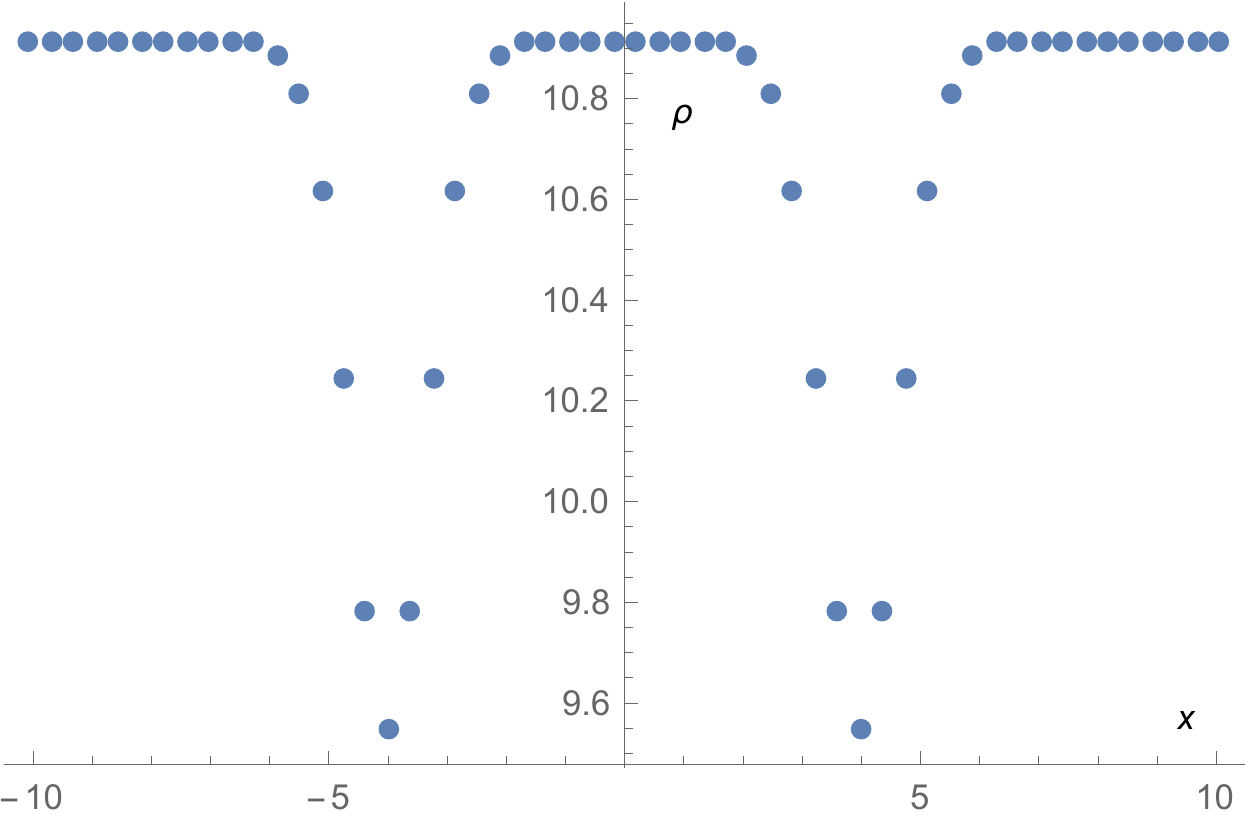}
\end{center}
\caption{Condensate and charge density of the double soliton as a function of $x$ at $\mu=9$.}\label{fig7}
\end{figure}

For the double parallel dark soliton solution, the seed configurations are chosen as $\phi(x,z)=a[\tanh(x+b)-1-\tanh(x-b)]$ and $\varphi(x,z)=c$, where $a,b,c$ are constants. In this situation, the fields can have periodic symmetry in the $x$ direction(We have checked that a solution also exists for non-periodic boundary conditions in the $x$ direction. See also the following triple dark soliton case.). So we can choose the Fourier pseudo-spectral method as the Chebyshev collocation points are sparse around $x=0$ in the $x$ direction, and still use the Chebyshev pseudo-spectral method in the $z$ direction. Setting $a=6$, $b=4$, $c=1$ for chemical potential $\mu=9$, we obtain the solution of $\phi(x,z)$ and $\varphi(x,z)$ as shown in Fig.\ref{fig6} and the condensate and charge density in Fig.\ref{fig7}. When $b$ is big, in a wide range of $a$ and $c$, there is always a double soliton solution. But when $b$ is small enough, e.g. $b=0.5$, the result reduces to a homogeneous and isotropic superfluid solution.

\begin{figure}
\begin{center}
\includegraphics[scale=0.7]{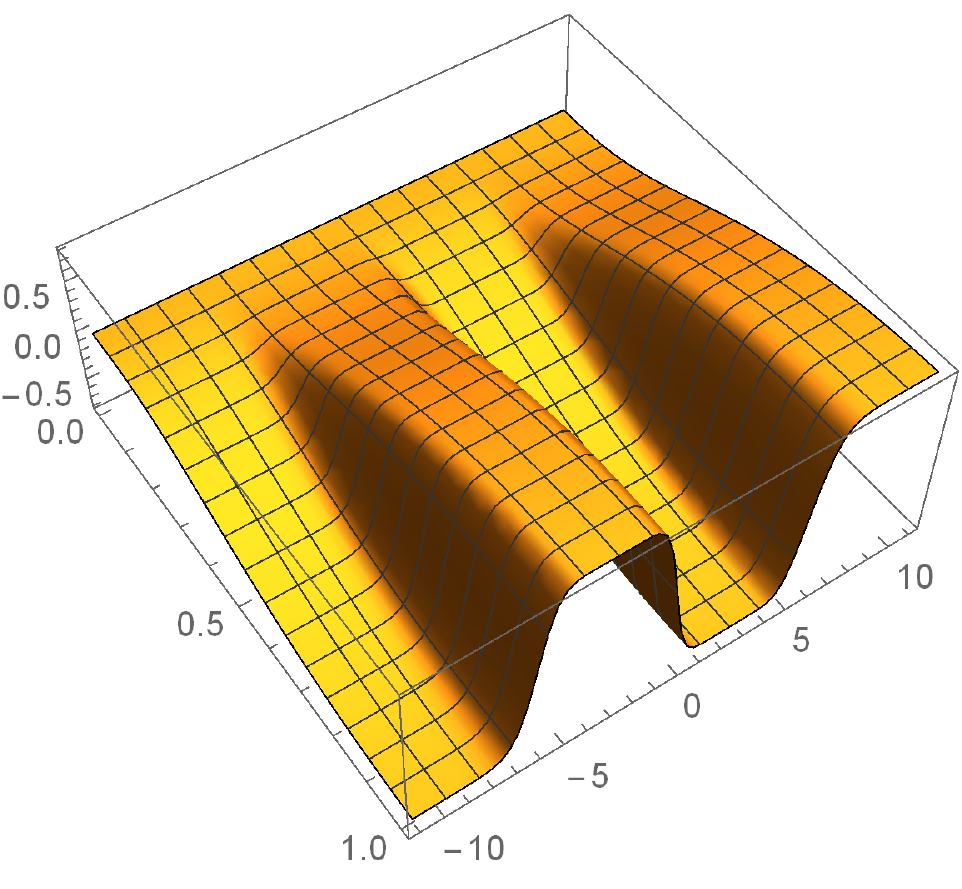}
\includegraphics[scale=0.7]{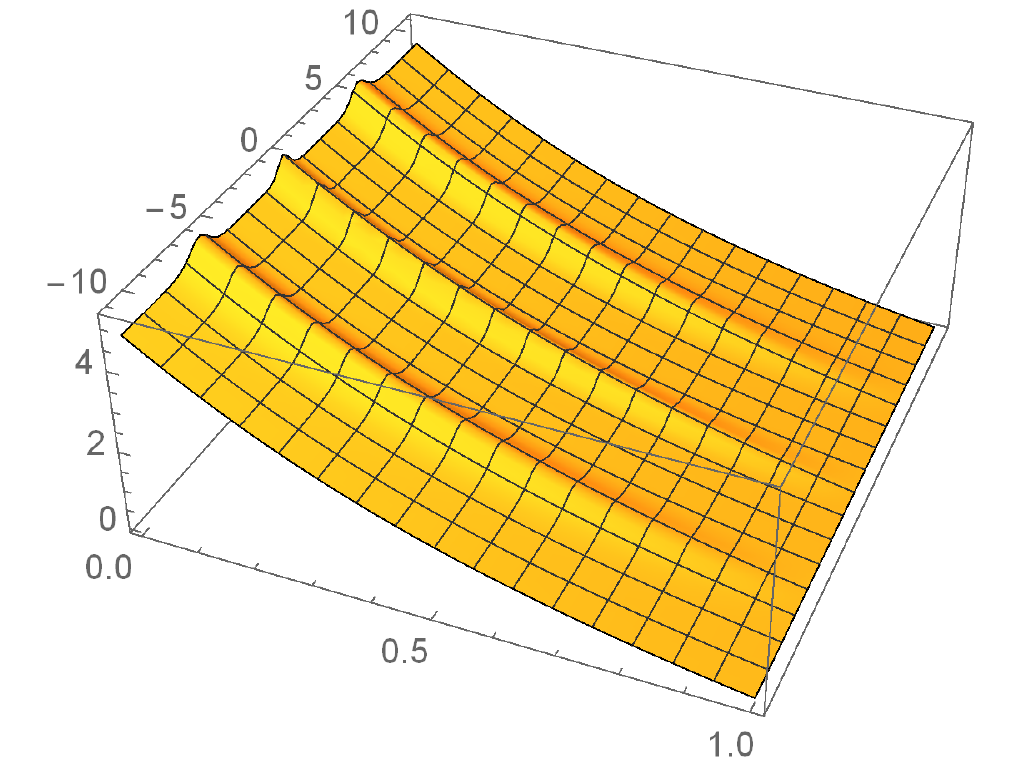}
\end{center}
\caption{The bulk configurations of the triple soliton fields $\phi(x,z),\varphi(x,z)$ from up to down respectively for $\mu=9$.}\label{fig8}
\end{figure}

\begin{figure}
\begin{center}
\includegraphics[scale=0.5]{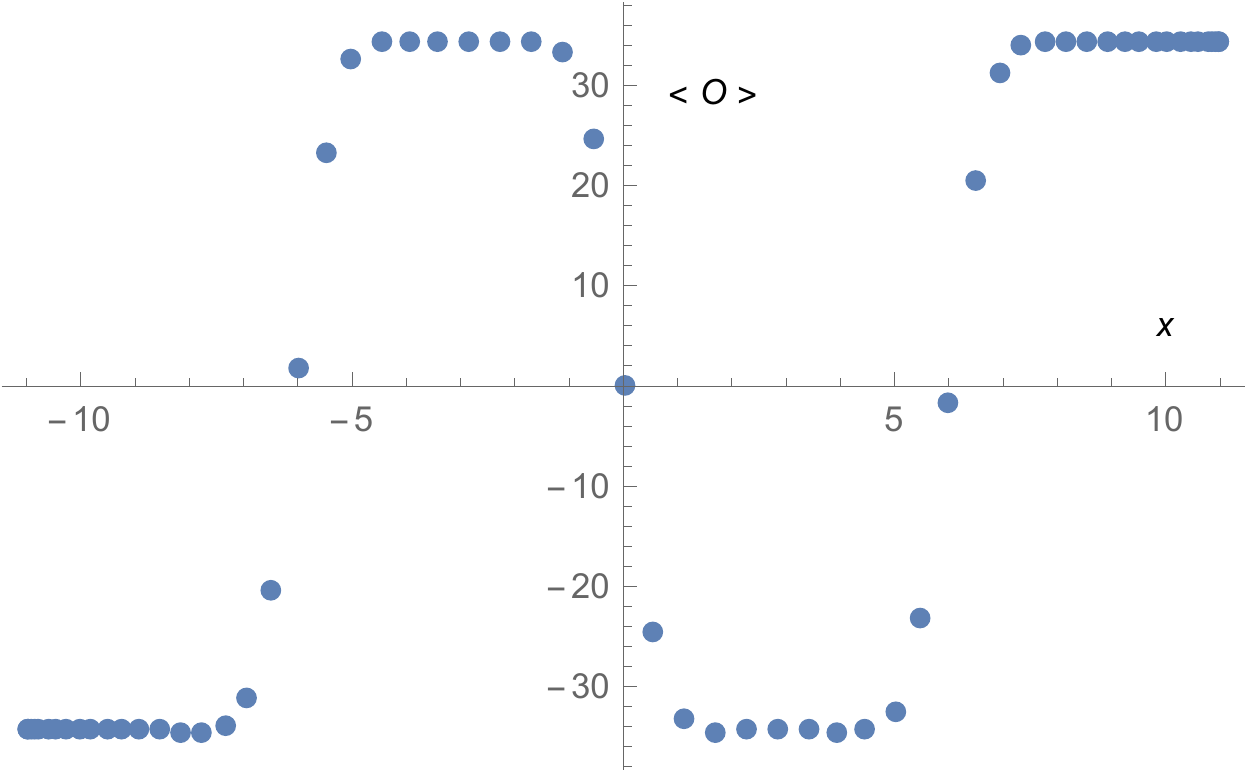}
\includegraphics[scale=0.5]{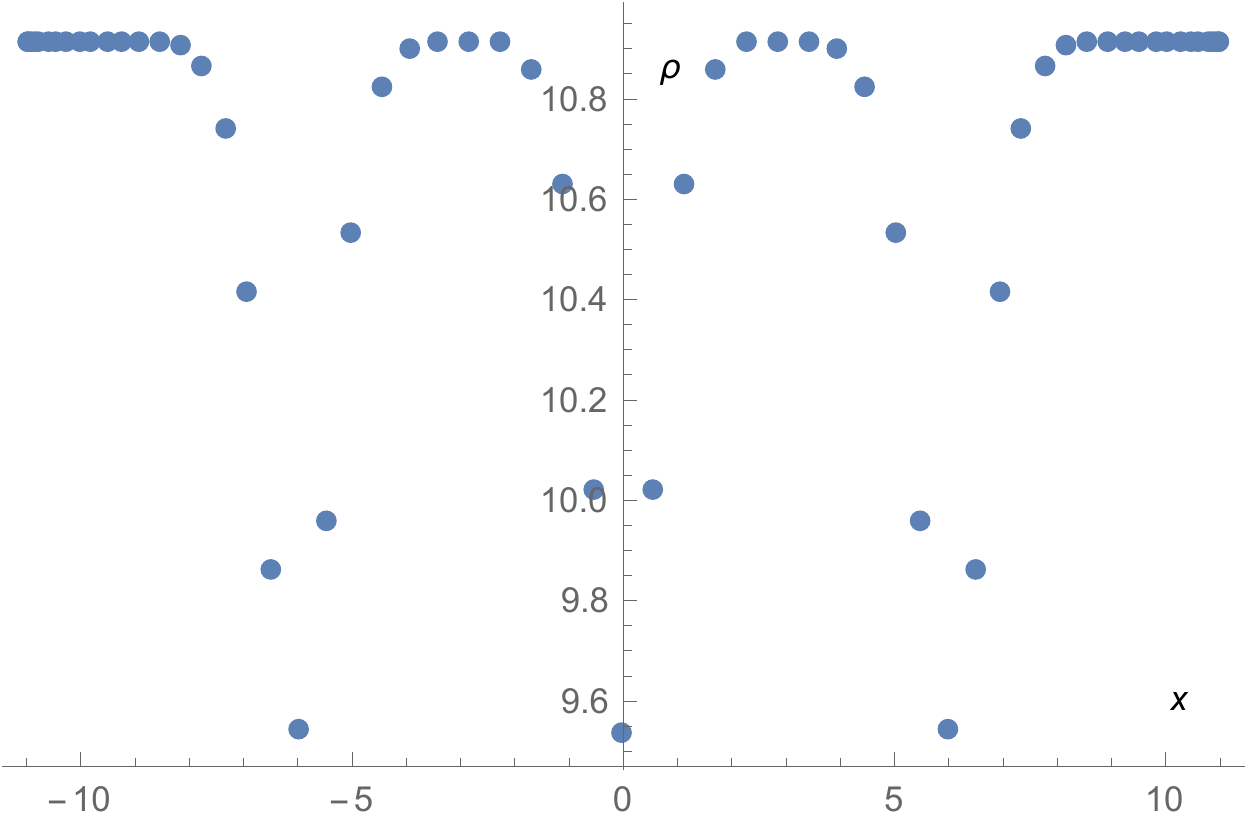}
\end{center}
\caption{Condensate and charge density of the triple soliton as a function of $x$ at $\mu=9$.}\label{fig9}
\end{figure}

For the triple parallel dark soliton solution, the seed configurations are chosen as $\phi(x,z)=a[\tanh(x+b)-\tanh(x)+\tanh(x-b)]$ and $\varphi(x,z)=c$, where $a,b,c$ are constants. In this case, the $x$ direction cannot be periodic and we then use the Chebyshev pseudo-spectral method in both the $x$ and $z$ directions. Setting $a=6$, $b=6$, $c=1$ for chemical potential $\mu=9$, we obtain the solution of $\phi(x,z)$ and $\varphi(x,z)$ as shown in Fig.\ref{fig8} and the condensate and charge density in Fig.\ref{fig9}. When $b$ is big, in a wide range of $a$ and $c$, there is always a triple soliton solution. But when $b$ is small enough, the result reduces to a double soliton solution.

\begin{figure}
\begin{center}
\includegraphics[scale=0.5]{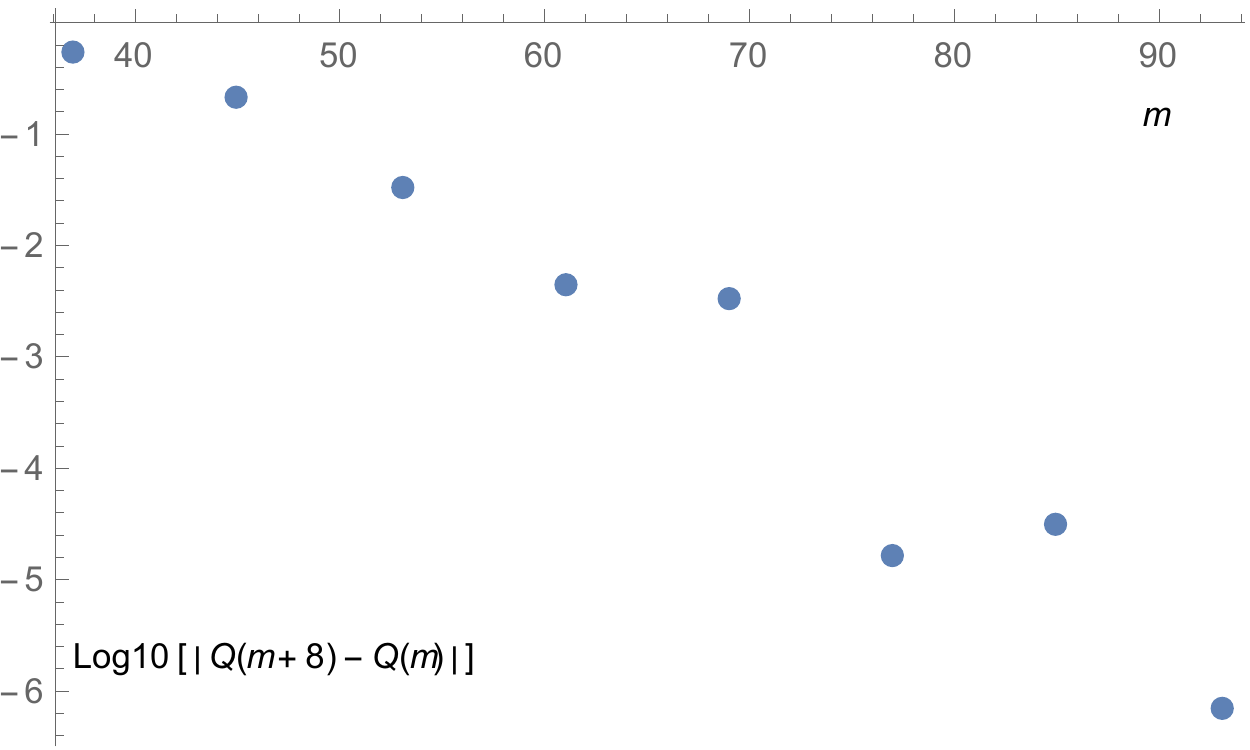}
\end{center}
\caption{$m$ is collocation points in $x$ direction and $Q(m)$ is the total charge calculated for $m$ collocation points(see, Fig.\ref{fig9}). Quantity $|Q(m+8)-Q(8)|$ has a logarithm decrease with respect to $m$ which denotes a very high precision of the numerical results. }\label{wucha}
\end{figure}
We have also investigated the convergence of our numerical solutions of multiple dark solitons, as shown in Fig.\ref{wucha}. It can be seen that these solutions satisfy the equations of motion at a very high precision, but it is impossible at the numerical level to determine whether a solution is exact or not. Nevertheless, the multiple dark soliton solutions constructed above are at least very well approximate solutions of the holographic superfluids, which may play an important role in superfluid physics at finite temperature.

For the periodic (double dark soliton) case, the solution can actually be viewed as an infinite sequence of soliton-anti-soliton pairs. It is easy to see by symmetry arguments that there should be an exact solution if the (anti-)solitons are equally spaced, since the forces of interaction on an (anti-)soliton are balanced, similar to the GPE case. But there is no obvious reason for the existence of an exact solution in the unequally spaced case or under non-periodic boundary conditions. So it remains an interesting open question to see the nature of the multiple dark soliton solutions we have constructed numerically. To end this section, we would like to point out that it is reasonable to conjecture that one can construct multiple soliton solutions with arbitrary number of single soliton and with different distances between them.

\section{inhomogeneous structures: vortex}
\label{vortex}

The holographic superconductor vortex solutions  (the sources for the flows are turned on) are studied in Refs.\cite{montull2009holographic,albash2009vortex,maeda2010vortex,rozali2012holographic,haiqing2015} for different $m^{2}$ models in the  Schwarzschild coordinate. The holographic superfluid vortex solutions (the sources for the flows are turned off) are studied in Refs.\cite{keranen2010inhomogeneous,keranen2011solitons} for different $m^{2}$ models in the Schwarzschild coordinate. Here we reproduce the superfluid vortex solutions for $m^{2}=0$ model in the infalling Eddington coordinates. Then we study the charge density and healing length of vortex solutions with different winding numbers at different chemical potential.

Single vortex solutions have cylindrical symmetry. When we rewrite the metric as
\begin{eqnarray}
ds^{2}=\frac{1}{z^{2}}(-f(z)dt^{2}-2dt dz+dr^{2}+r^{2}d\theta^{2}),
\end{eqnarray}
all the gauge invariant fields will be independent of $\theta$ coordinate. Then we can rewrite the complex scalar field as $\Psi(r,\theta,z)=\phi(r,z)e^{i\tilde{\varphi}(r,\theta,z)}$ where $\phi$ is gauge invariant. What is more, the currents in Eq.(5) are also independent of $\theta$. The currents in $r$ and $z$ directions should be zero. Thus we have three constraint equations as
\begin{equation}
A_{r}(r,\theta,z)-\partial_{r}\tilde{\varphi}(r,\theta,z)=0,
\end{equation}
\begin{equation}
A_{t}(r,\theta,z)+f(z)\partial_{z}\tilde{\varphi}(r,\theta,z)=0,
\end{equation}
\begin{equation}
\partial_{\theta}(A_{\theta}(r,\theta,z)-\partial_{\theta}\tilde{\varphi}(r,\theta,z))=0.
\end{equation}
 In order to have a single valued function $\Psi$, its phase term has to be $\tilde{\varphi}(r,\theta,z)=n \theta+\varphi(r,z)$.  Constant $n$ is an integer which denotes the quantized winding number of vortex. As a result, the three constraint equations become
\begin{equation}
A_{r}(r,z)-\partial_{r}\varphi(r,z)=0,
\end{equation}
\begin{equation}
A_{t}(r,z)+f(z)\partial_{z}\varphi(r,z)=0,
\end{equation}
\begin{equation}
\partial_{\theta}A_{\theta}(r,\theta,z)=0.
\end{equation}
By using these constraint equations, the equations of motion will be dramatically simplified,
\begin{eqnarray}
&\,&r^{2}zf\partial_{z}^{2}\phi+r^{2}z\partial_{r}^{2}\phi-(3-f)r^{2}\partial_{z}\phi+rz\partial_{r}\phi\nonumber\\
&\,&+r^{2}zf\phi(\partial_{z}\varphi)^{2}-z\phi(A_{\theta}-n)^{2}=0, \label{veom1}
\end{eqnarray}
\begin{eqnarray}
&\,&z^{2}(\partial_{r}\partial_{z}\varphi+r\partial_{r}^{2}\partial_{z}\varphi-6rz\partial_{z}\varphi-6rz^{2}\partial_{z}^{2}\varphi\nonumber\\
&\,&+rf\partial_{z}^{3}\varphi)-2r\phi^{2}\partial_{z}\varphi=0, \label{veom2}
\end{eqnarray}
\begin{eqnarray}
&\,&rz^{2}(\partial_{r}^{2}A_{\theta}+f\partial_{z}^{2}A_{\theta}-3z^{2}\partial_{z}A_{\theta})-z^{2}\partial_{r}A_{\theta}\nonumber\\
&\,&-2r(A_{\theta}-n)\phi^{2}=0. \label{veom3}
\end{eqnarray}

Now all the above field functions are independent of $\theta$. So we only need to consider the system in a box of size $1\times L$ in the $z$ and $r$ directions respectively. The boundary conditions are introduced as follows. At $z=0$, $\phi(z=0)=A_{\theta}=0$ (sources turned off) and $\partial_{z}\varphi(z=0)=-\mu$ (choose a chemical potential). At $z=1$, we put regular boundary conditions for $\phi$ and $A_{\theta}$ functions, and $\varphi(z=1)=0$ (introduced as Eq.(33) with the same argument as before). At $r=0$, we put regular boundary conditions for $\varphi$ and $A_{\theta}$ functions, and $\phi(r=0)=0$(the condensate at vortex core is zero). At $r=L$, we will introduce Neumann boundary conditions for all the fields $\phi$, $\varphi$ and $A_{\theta}$. With the equations of motion and boundary conditions set up, we can use relaxation methods to find solutions. First, we choose seed configurations $\phi(r,z)=a \,\,tanh(b\,r)$, $\varphi(r,z)=c$ and $A_{\theta}(r,z)=d$ where $a,b,c,d$ are constants. Then relax these configurations towards the solutions.

\begin{figure}
\begin{center}
\includegraphics[scale=0.6]{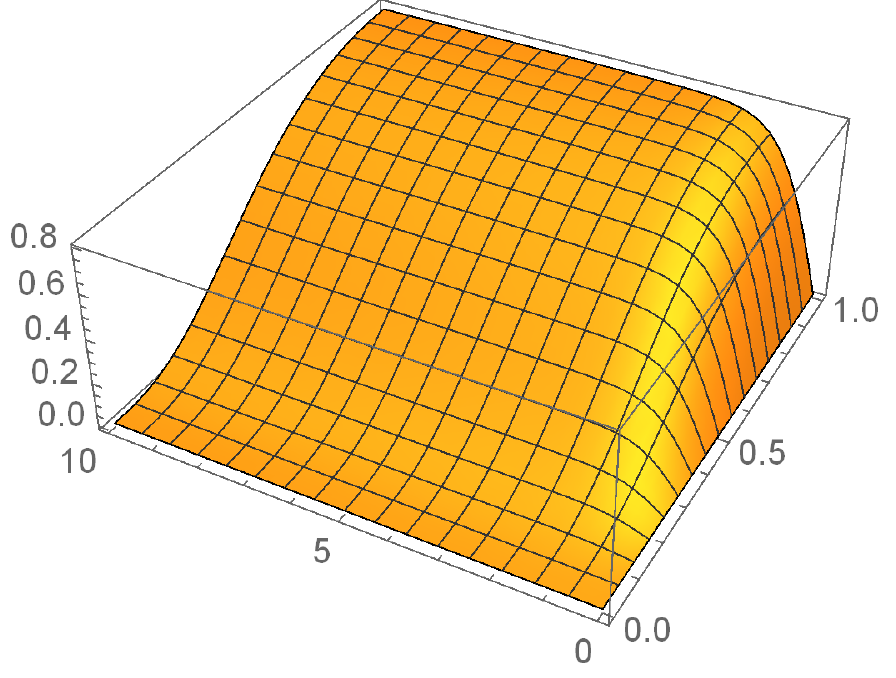}
\includegraphics[scale=0.6]{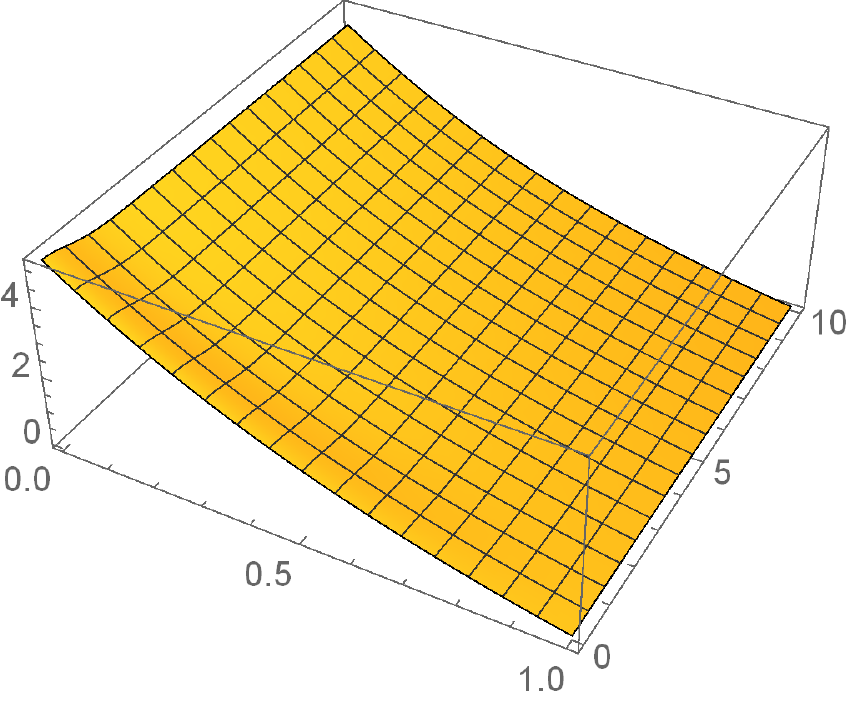}
\includegraphics[scale=0.6]{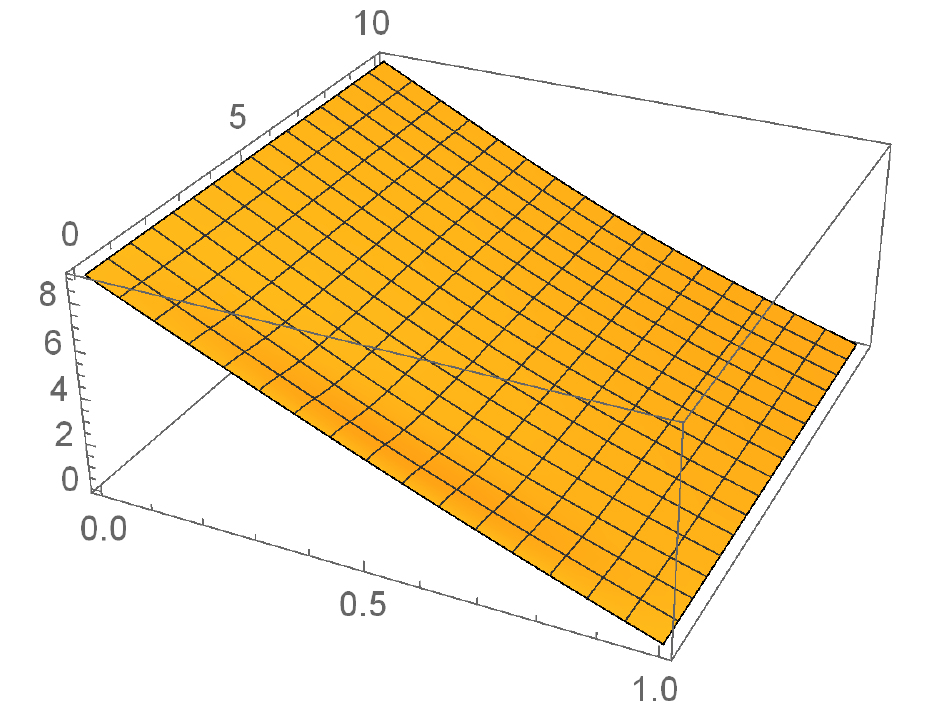}
\includegraphics[scale=0.6]{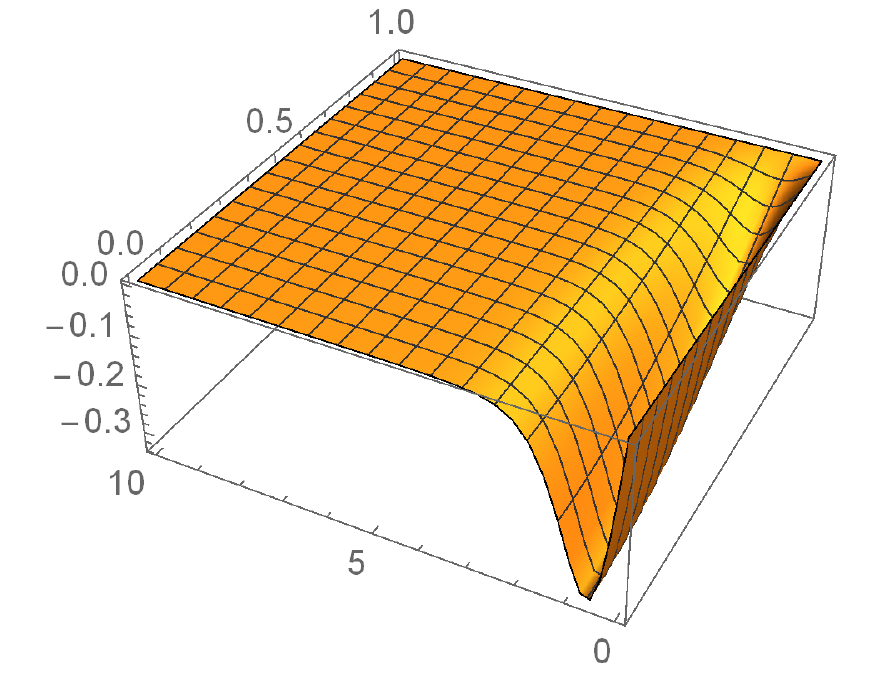}
\includegraphics[scale=0.6]{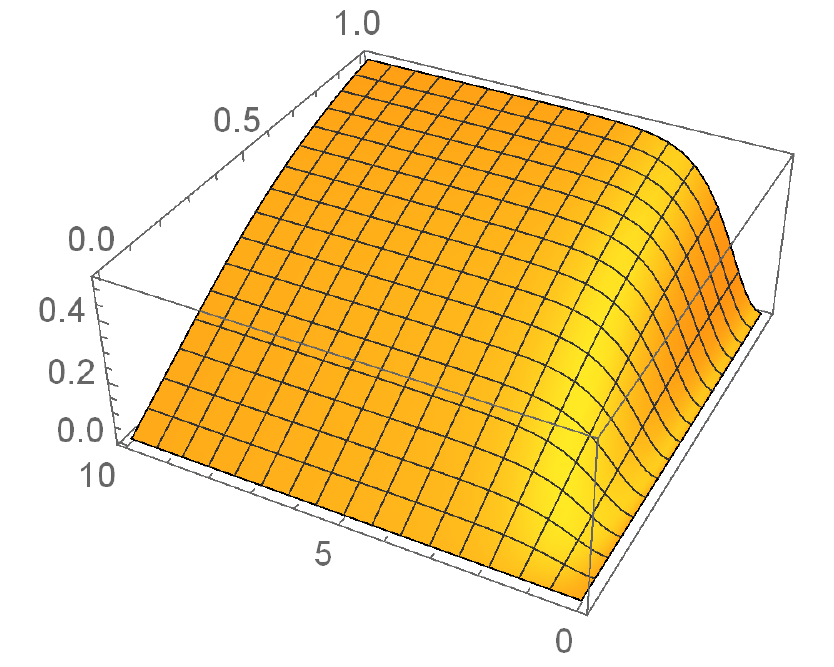}
\end{center}
\caption{The bulk configurations of the fields $\phi(r,z),\varphi(r,z),A_{t}(r,z),A_{r}(r,z),A_{\theta}(r,z)$ from up to down respectively for $\mu=9$.}\label{fig10}
\end{figure}

\begin{figure}
\begin{center}
\includegraphics[scale=0.5]{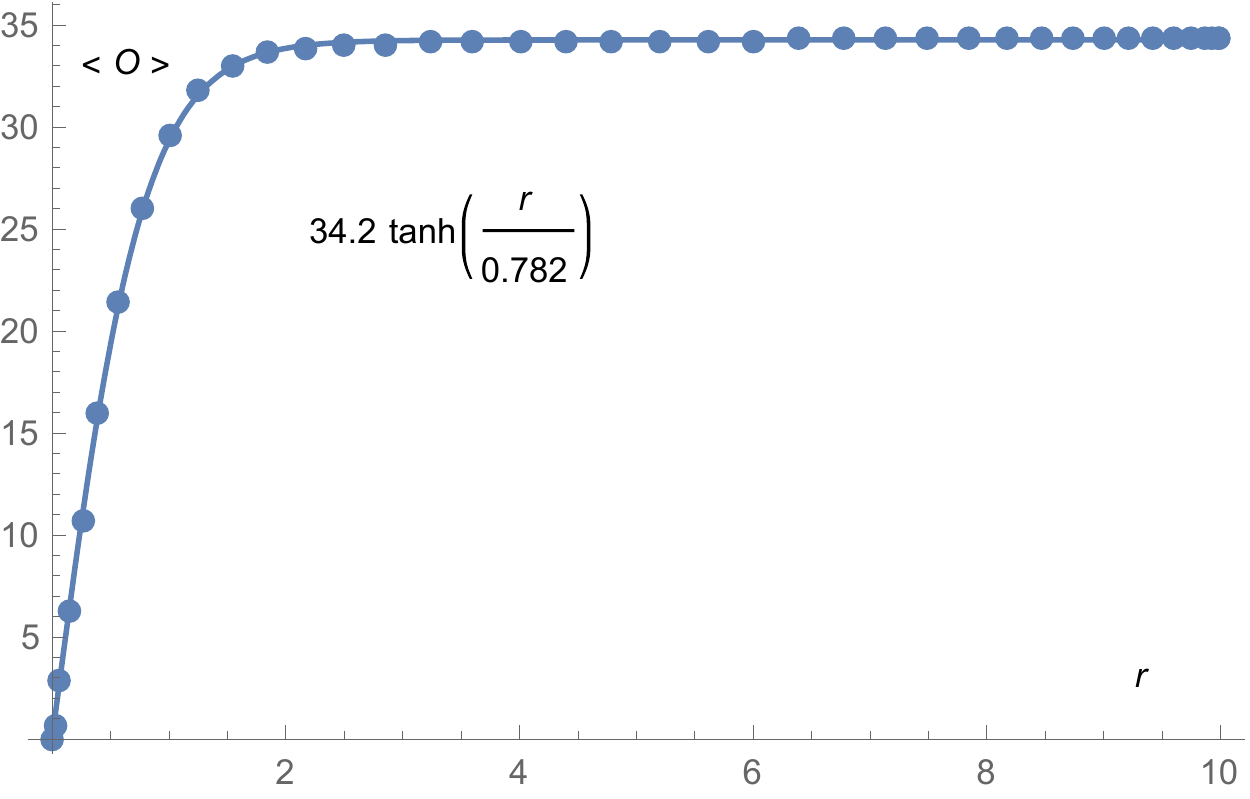}
\includegraphics[scale=0.5]{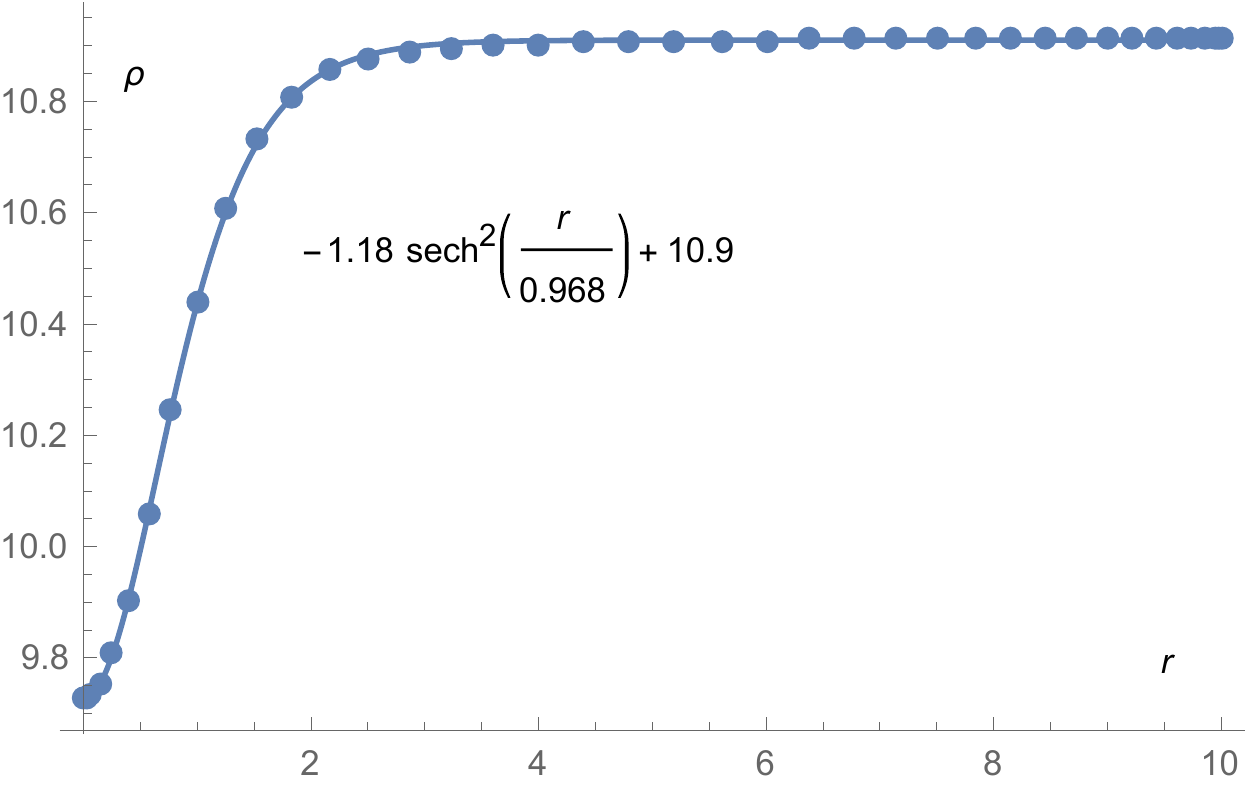}
\end{center}
\caption{Condensate and charge density of vortex as functions of $r$ at $\mu=9$. The solid lines are fitting graphes of respective functions.}\label{fig11}
\end{figure}

For vortex with winding number $n=1$ and chemical potential $\mu=9$, the numerical results of the bulk fields configurations are showed in Fig.\ref{fig10}. Condensate and charge density as functions of $r$ are showed in Fig.\ref{fig11}.
The condensate is well fitted by function
\begin{equation}
\langle O \rangle (r)=34.2\tanh(\frac{r}{0.782}),
\end{equation}
where $34.2$ is exactly the condensate of homogeneous and isotropic solution with the same chemical potential and $0.782$ should be thought of as healing length of the vortices. The charge density is well fitted by function
\begin{equation}\label{rhosech}
\rho(r)=-1.18\,sech^{2}(\frac{x}{0.968})+10.9,
\end{equation}
where $10.9$ is exactly the charge density of homogeneous and isotropic solution with the same chemical potential, $0.968$ should also be thought of as healing length of the vortex and $1.18$ is the charge density depletion. Thus there are two healing lengths (one for condensate and the other for charge density) for holographic vortex. From Table.\ref{table1} and Table.\ref{table2}, comparing the results with that of the holographic soliton solutions, holographic vortices have the same forms of fitting functions and two healing lengths. But the healing lengths are bigger and the charge density depletion is smaller at the same chemical potential. Comparing the results with that of the vortex solutions of the GPE, the difference are that the charge density in the vortex core is zero and there are two healing lengthes (but one for the vortex core and the other for the vortex tail) for the GPE solutions.

\subsection{vortex with winding number $n=1$ at different chemical potential}

For vortices with winding number $n=1$ at different chemical potential, the parameters which characterise the vortices are listed in Table.\ref{table2}.
\begin{table}
\begin{center}
\begin{tabular}{|c|c|c|c|c|c|c|c|}
  \hline
  \,\,\,\,$\mu$\,\,\,\, & \,\,$\langle O \rangle_{max}$ \,\,& \,\,\,$\rho_{max}$\,\,\, & \,\,\,\,$\delta\rho$ \,\,\,\,&\,\,$\delta\rho/\rho_{max}$ \,\,& \,\,\,\,$\epsilon_{c}$ \,\,\,\,& \,\,\,\,$\epsilon_{\rho}$\,\,\,\, \\
   \hline
  8 & 15.0 & 8.51 & 0.41 &0.048 & 1.515 & 1.704 \\
  \hline
  9 & 34.2 & 10.9 & 1.18 &0.108 & 0.782 & 0.968 \\
  \hline
  10 & 54.5 & 13.6 & 1.84 &0.136 & 0.591 & 0.762\\
  \hline
  11 & 77.9 & 16.5 & 2.49 &0.151 & 0.493 & 0.649 \\
  \hline
  12 & 105.4 & 19.7 & 3.17 &0.161 & 0.431 & 0.573 \\
  \hline
  13 & 137.5 & 23.1 & 3.87 &0.167 & 0.386 & 0.517 \\
  \hline
  14 & 174.7 & 26.9 & 4.61 & 0.172 &0.350 & 0.473 \\
  \hline
  15 & 217.3 & 30.9 & 5.37 & 0.174 &0.321 & 0.436 \\
  \hline
  16 & 266.0 & 35.1 & 6.15 & 0.175 &0.296 & 0.405 \\
  \hline
  17 & 321.2 & 39.7 & 6.95 & 0.175 &0.274 & 0.378 \\
  \hline
  18 & 383.2 & 44.5 & 7.78 & 0.175 &0.255 & 0.354 \\
  \hline
  19 & 452.6 & 49.6 & 8.63 & 0.174 &0.239 & 0.333 \\
  \hline
  20 & 530.0 & 54.9 & 9.53 & 0.174 &0.225 & 0.315 \\
  \hline
\end{tabular}
\end{center}
\caption{The relation between chemical potential $\mu$ and max condensate $\langle O \rangle_{max}$, max charge density $\rho_{max}$, charge density depletion $\delta \rho$, healing length $\epsilon_{c}$ for condensate, healing length $\epsilon_{\rho}$ for charge density at winding number $n=1$.\label{table2}}
\end{table}
As we increase the chemical potential(lower the temperature), (1) the condensate increases dramatically, (2) the depletion of charge density increases and the two healing lengths (width of the vortices) decreases, as a result, the vortices become thinner and smaller, (3) $\delta\rho/\rho_{max}$ increases fast near the critical point but slows down later, this ratio is small for all $\mu$ which means the depletion of vortices is shallow in the charge density and it is not likely to increase to 1 (the charge density at the vortex core is not likely to be 0). Actually, it seems to set down at $0.175$.

\subsection{vortex with different winding number at certain chemical potential}

\begin{figure}
\begin{center}
\includegraphics[scale=0.5]{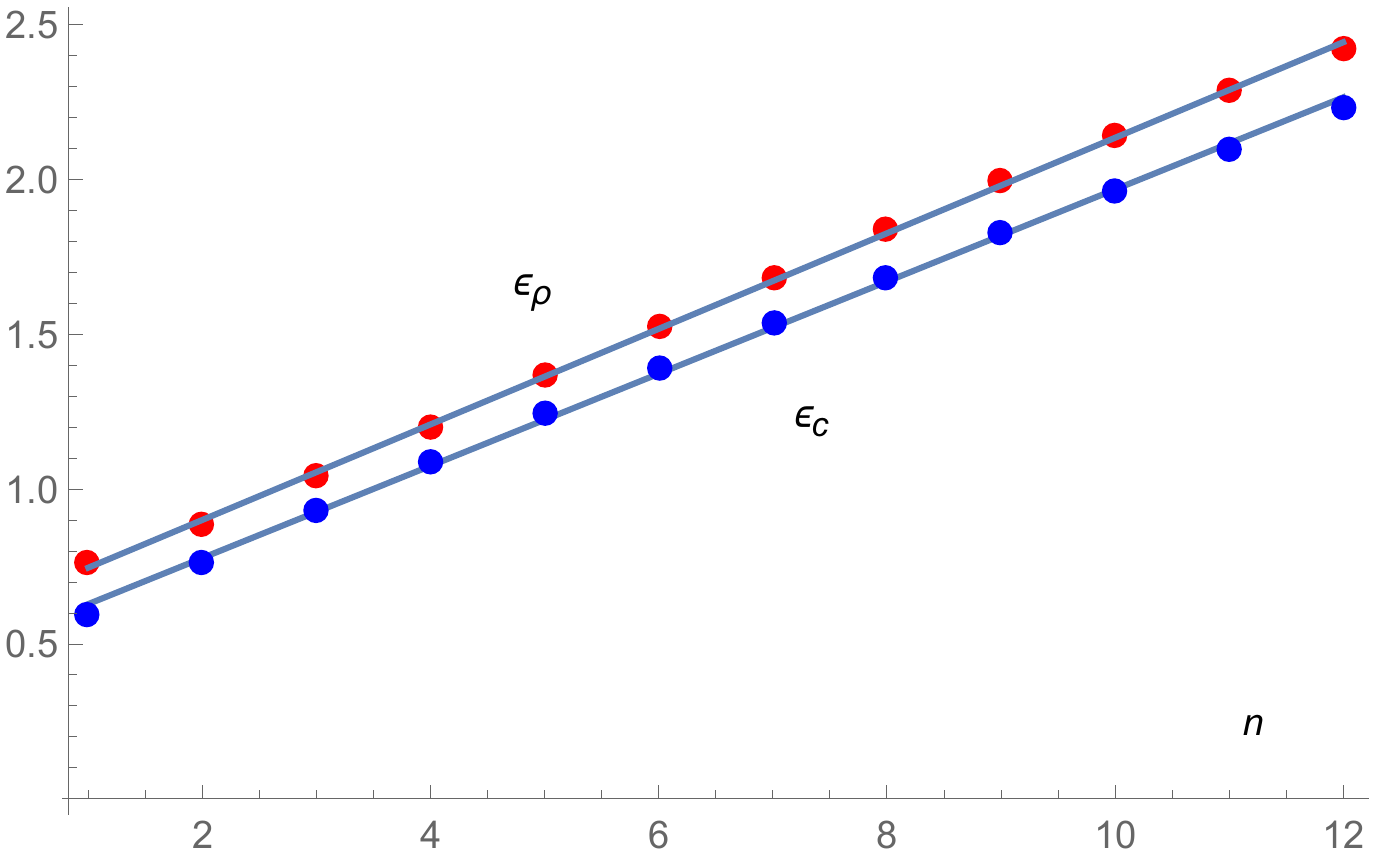}
\end{center}
\caption{The relation between winding number $n$ and healing length $\epsilon_{c}$ for condensate, healing length $\epsilon_{\rho}$ for charge density which are obtained by Eqs.(40) and (41) at chemical potential $\mu=10$.}\label{fig12}
\end{figure}

\begin{figure}
\begin{center}
\includegraphics[scale=0.5]{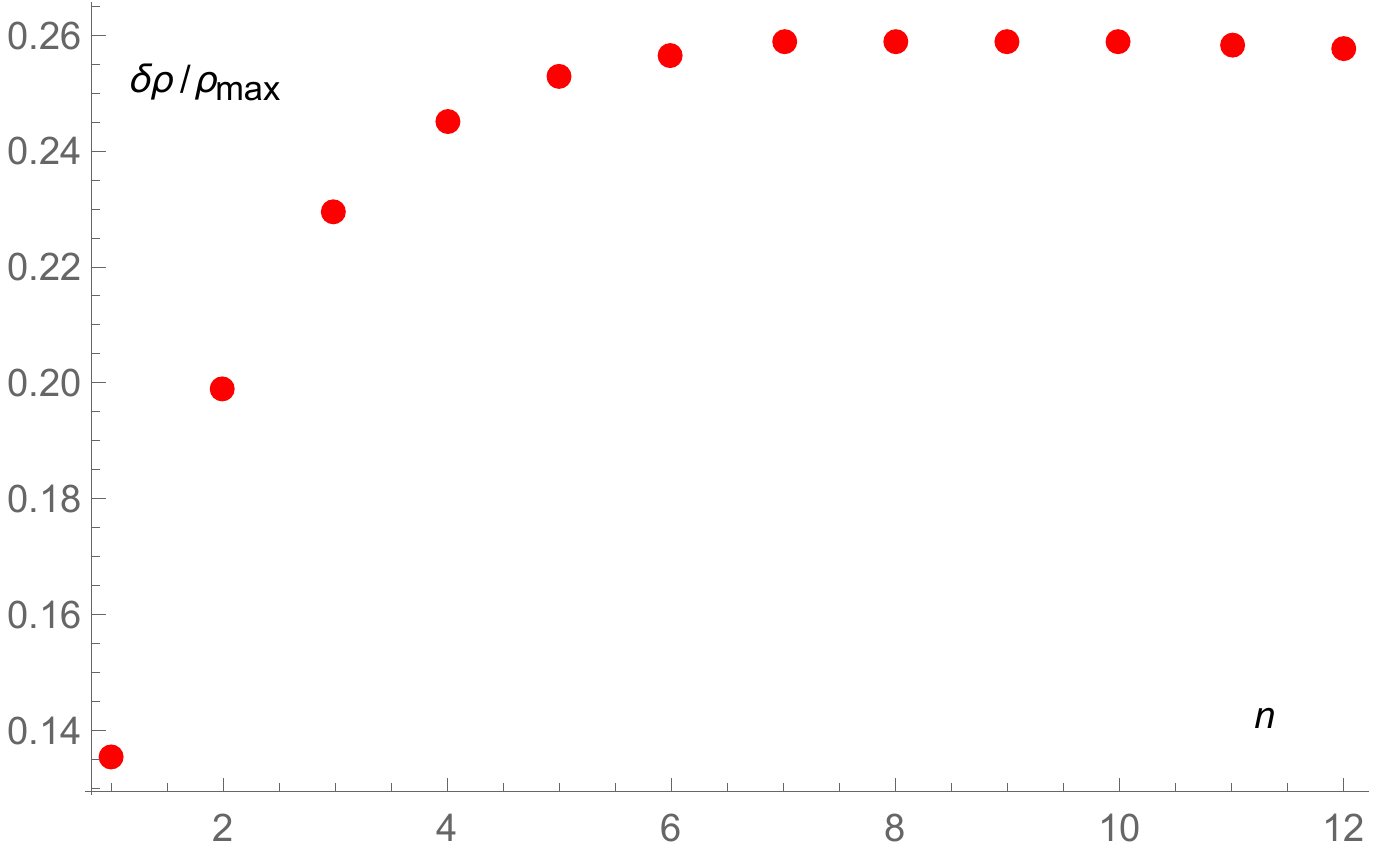}
\end{center}
\caption{The relation between winding number $n$ and the ratio of charge density depletion to max charge density $\delta \rho/\rho_{max}$ at chemical potential $\mu=10$.}\label{fig13}
\end{figure}

For vortices with different winding number $n$ at chemical potential $\mu=10$, the condensates are well fitted by function
\begin{equation}
\langle O \rangle (r)=\langle O \rangle_{max}\,\, tanh^{n}(\frac{r}{\epsilon_{c}}),
\end{equation}
where $\langle O \rangle_{max}$ is the max condensate of its profile and $\epsilon_{c}$ should be thought of as healing length of the vortices. The charge density is well fitted by function
\begin{equation}\label{rhotanh}
\rho(r)=\delta\rho \,\,tanh^{n+1}(\frac{r}{\epsilon_{\rho}})+\rho_{0},
\end{equation}
where $\rho_{max}=\rho_{0}+\delta\rho$ is the max charge density of its profile and $\epsilon_{\rho}$ should also be thought of as healing length of the vortices and $\delta\rho$ is the charge density depletion. It is easy to check that when $n=1$, the fitting formulas for charge density (Eq.(\ref{rhosech}) and Eq.(\ref{rhotanh})) are equivalent.
The two healing lengthes and the ratio of charge density depletion to max charge density are plotted in Fig.\ref{fig12} and in Fig.\ref{fig13}. These two Eqs.(40) and (41) are well fitting at least for $n\leq12$.

As we increase the winding number, (1) the two healing lengthes increase linearly, (2) the ratio of charge density depletion to max charge density $\delta \rho/\rho_{max}$ (or say the charge density depletion $\delta \rho$ as the max charge density is that of homogeneous and isotropic superfluid) increases fast at small winding number but sets down at a certain value (around $0.258-0.259$) at large winding number. As a result, the vortices will become fatter (bigger, but not deeper) if we continue to increase the winding number.

\begin{figure}
\begin{center}
\includegraphics[scale=0.5]{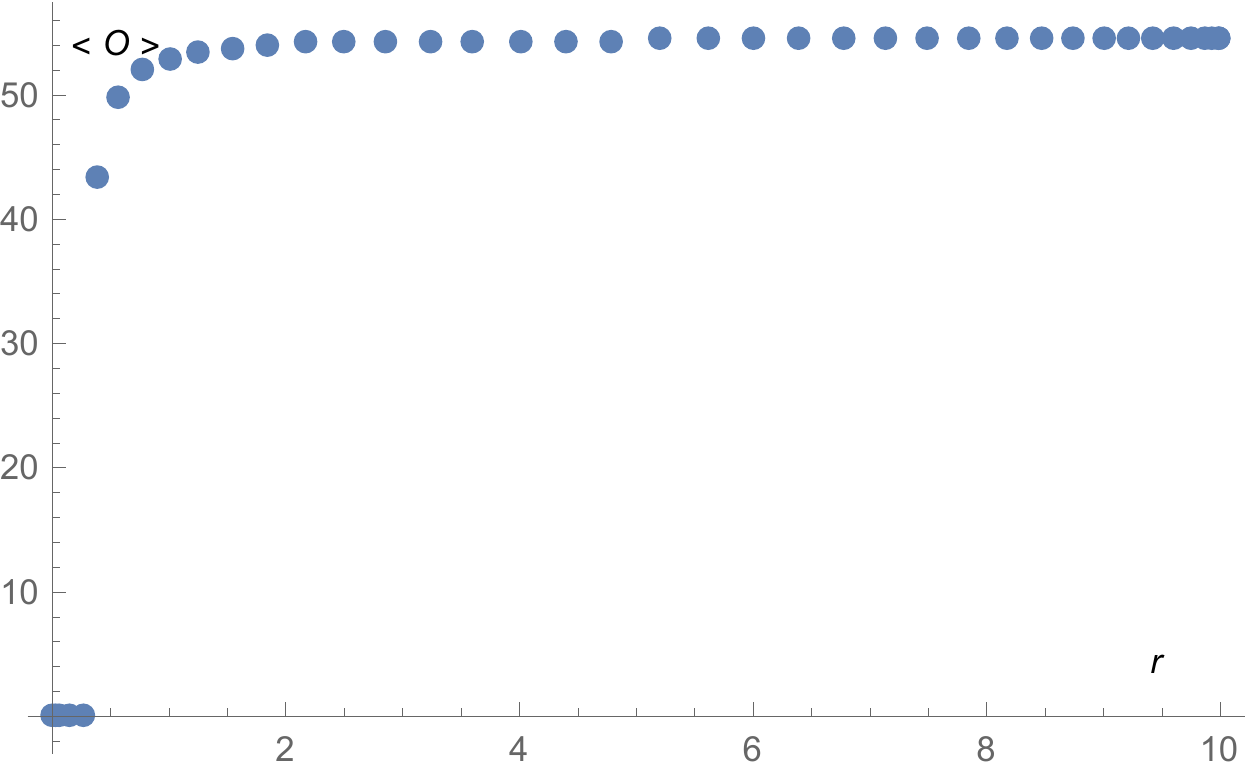}
\includegraphics[scale=0.5]{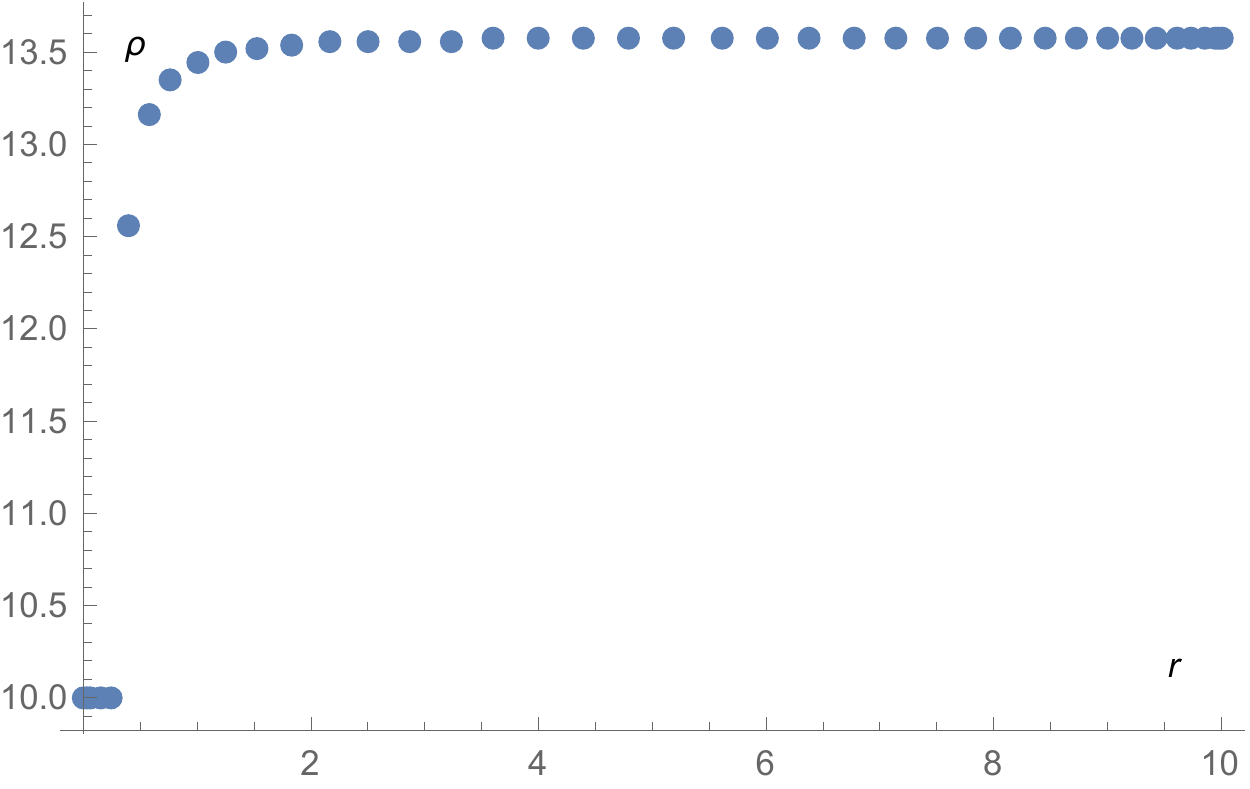}
\end{center}
\caption{Condensate and charge density of vortex as a function of $r$ at $\mu=10$.}\label{fig14}
\end{figure}

To understand this phenomenon, we make such a scaling
\begin{eqnarray}
&\,&r=n\tilde{r},\nonumber\\
&\,&A_{\theta}=n\tilde{A_{\theta}},\label{rescaling}
\end{eqnarray}
and plug them into Eqs.(\ref{veom1}-\ref{veom3}). Setting $n\rightarrow\infty$, the equations of motion become
\begin{eqnarray}
&\,&\tilde{r}^{2}(zf\partial_{z}^{2}\phi-(3-f)\partial_{z}\phi+zf\phi(\partial_{z}\varphi)^{2})\nonumber\\
&\,&-z\phi(\tilde{A}_{\theta}-1)^{2}=0,
\end{eqnarray}
\begin{eqnarray}
z^{2}(f\partial_{z}^{3}\varphi-6z\partial_{z}\varphi-6z^{2}\partial_{z}^{2}\varphi)-2\phi^{2}\partial_{z}\varphi=0,
\end{eqnarray}
\begin{eqnarray}
z^{2}(f\partial_{z}^{2}\tilde{A}_{\theta}-3z^{2}\partial_{z}\tilde{A}_{\theta})-2(\tilde{A}_{\theta}-1)\phi^{2}=0.
\end{eqnarray}
Surprisingly, these equations are independent of winding number $n$. As a result, the system has a scaling symmetry Eq.(\ref{rescaling}) for large $n$. The scaling symmetry includes linear change of $r$ coordinate and leaves $z$ coordinate unchanged with respect to $n$, so the two healing lengthes which are in $r$ direction increase linearly and the charge density depletion which is independent of $r$ coordinate maintains the same value with respect to large $n$. The condensate and charge density profiles of the numerical vortex solution of the above equations at chemical potential $\mu=10$ are shown in Fig.\ref{fig14}. The ratio of charge density depletion to max charge density is $\delta \rho/\rho_{max}=3.6/13.6=0.265$ . Comparing the result with the max charge density depletion ratio in Fig.\ref{fig13}, one can see that they are consistent.

\section{Summary and Discussion}

For the $m^{2}=0$ Abelian Higgs Model in AdS$_{4}$, in the infalling Eddington coordinates, we have numerically solved the equations of motion and investigated in detail three types of static holographic superfluid solutions: the homogeneous and isotropic superfluid solutions, the dark soliton solutions and the vortex solutions.

For the homogeneous and isotropic holographic superfluid solutions, we have calculated the sound speeds. When increasing the chemical potential, the square of sound speed approaches to $1/2$. This result is consistent with that in Refs. \cite{herzog2009sound,yarom2009fourth}. It is argued that the sound speed is $1/2$ for conformal fluids at zero temperature besides some exceptions when the conformal dimension of condensate $\Delta<2$. Our $m^{2}=0$ case corresponds to the conformal dimension $\Delta_{+}=3$ which is larger than 2. So the sound speed approaches $1/2$ when we increase the chemical potential.

Breaking the translation symmetry in $x$ direction, we have found the single, double and triple dark soliton solutions. For the single soliton solutions, the charge density depletion and the two healing lengths (one for the condensate, the other for the charge density) are studied for different chemical potential by finding the well fitting functions. The results show that the dark soliton becomes thinner (with larger charge density depletion and smaller healing lengths) as we increase the chemical potential.

The multiple dark soliton solutions are constructed by superposing multiple single dark solitons as seed configurations. The distance between the solitons are controlled by the constant $b$ of the seed configurations. When the distance $b$ is large, the seed configurations relax to the solutions very quickly. When $b$ is very small, the corresponding double and triple soliton solutions seem not to exist, since the seed configurations of double soliton will lead to the homogeneous solution and the seed configurations of triple soliton will lead to a double soliton solution. Whether approximate or actually exact, these solutions have many potential applications. For example, in the investigation of superfluid turbulence, one needs an initial configuration which will lead to turbulence. In Refs.\cite{chesler2013holographic,Ewerz2014tua,du2014holographic,lan2016towards}, the initial configurations are constructed by putting vortices in the superfluid, which may introduce many unwanted modes when one try to make the periodic configuration. In contrast, the multiple dark soliton solutions can provide alternative and much cleaner initial configurations.

Considering the system with cylindrical symmetry, we found vortex solutions with different winding numbers at different chemical potential. For winding number $n=1$, if we increase the chemical potential, the depletion of charge density will increase and the two healing lengths will decrease. As a result, the vortices become thinner and smaller. At a certain chemical potential, if we increase the winding number, at first the vortices will become bigger and the charge density depletion will become larger, then the charge density depletion will settle down at a certain value and the vortices will become fatter. Very well fitting functions of the condensate and charge density are found. These features are controlled by a scaling symmetry as Eq.(\ref{rescaling}) for large $n$.

\hspace{2em}
\acknowledgments

We are grateful to Hongbao Zhang for his helpful discussions on the paper. W.L. is partially supported by NSFC with Grant Nos. 11235003, 11175019 and 11178007. Y.T. is partially supported by NSFC with Grant No.11475179 and by the Opening Project of Shanghai Key Laboratory of High Temperature Superconductors(14DZ2260700).

\section*{ Appendix: multiple soliton solutions for $m^{2}=-2$ model }

Following the discussions above, one can find that in $m^{2}=-2$ model, the equations of motion for the soliton solutions are
\begin{equation}
A_{x}(x,z)-\partial_{x}\varphi(x,z)=0,
\end{equation}
\begin{equation}
A_{t}(x,z)+f(z)\partial_{z}\varphi(x,z)=0,
\end{equation}
\begin{eqnarray}
f(\partial_{z}\varphi)^{2}\phi-z\phi-3z^{2}\partial_{z}\phi+f\partial_{z}^{2}\phi+\partial_{x}^{2}\phi=0,
\end{eqnarray}
\begin{eqnarray}
&\,&2\partial_{z}\varphi\phi^{2}+6z\partial_{z}\varphi+6z^{2}\partial_{z}^{2}\varphi\nonumber\\
&\,&-f\partial_{z}^{3}\varphi-\partial_{z}\partial_{x}^{2}\varphi=0.
\end{eqnarray}

For the conformal dimension $\Delta_{+}=2$ case at chemical potential $\mu=6$, the double and triple dark soliton solutions are presented below.
\begin{figure}
\begin{center}
\includegraphics[scale=0.7]{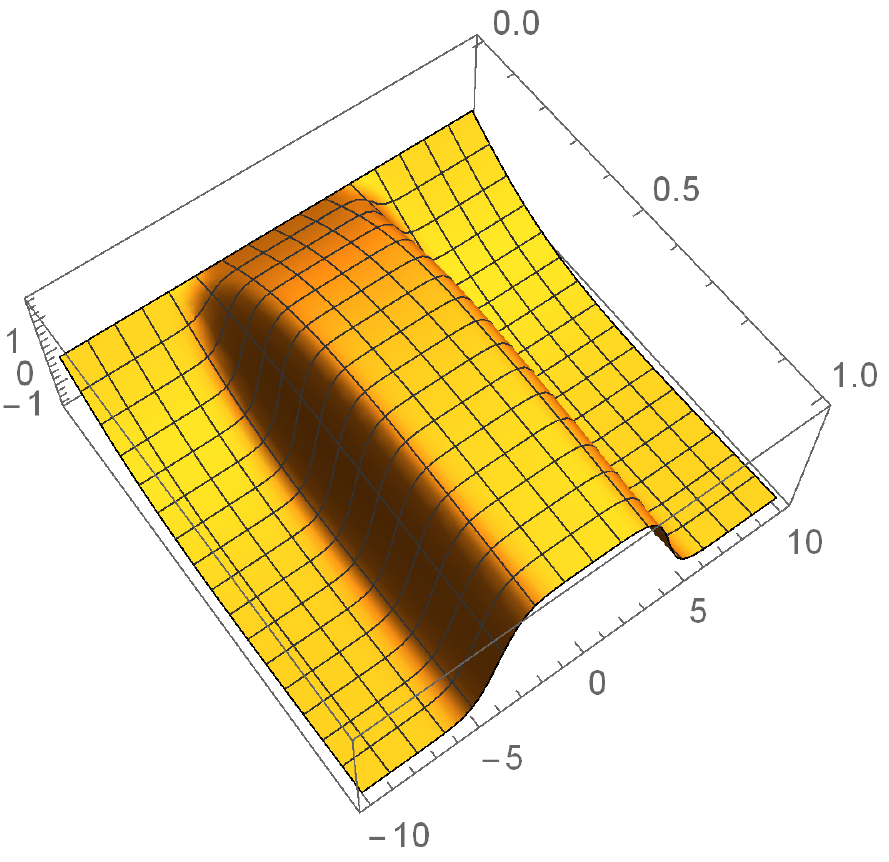}
\includegraphics[scale=0.7]{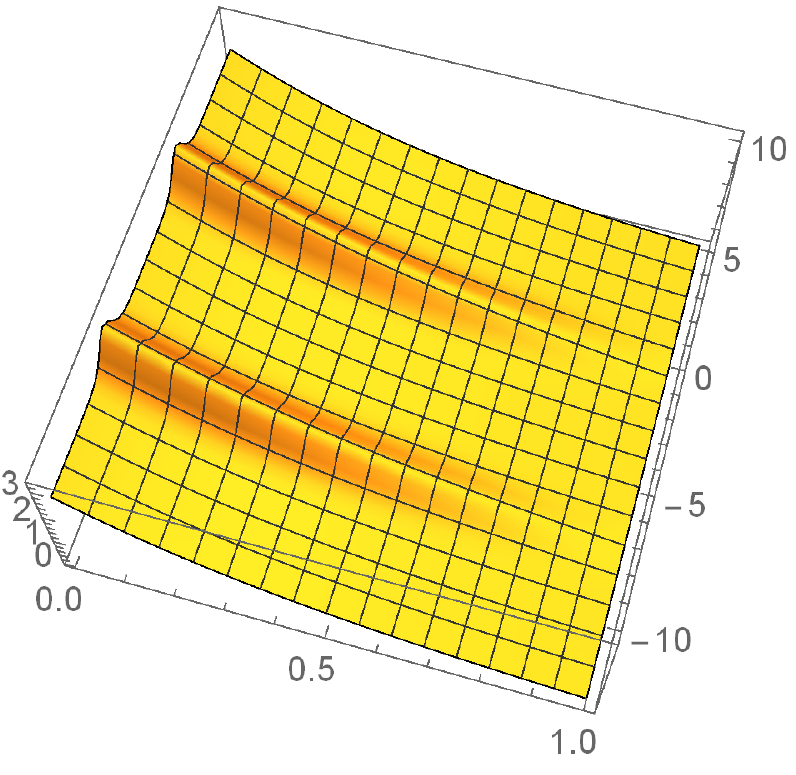}
\end{center}
\caption{The bulk configurations of the double soliton fields $\phi(x,z),\varphi(x,z)$ from up to down respectively for $\mu=6$.}
\end{figure}
\begin{figure}
\begin{center}
\includegraphics[scale=0.5]{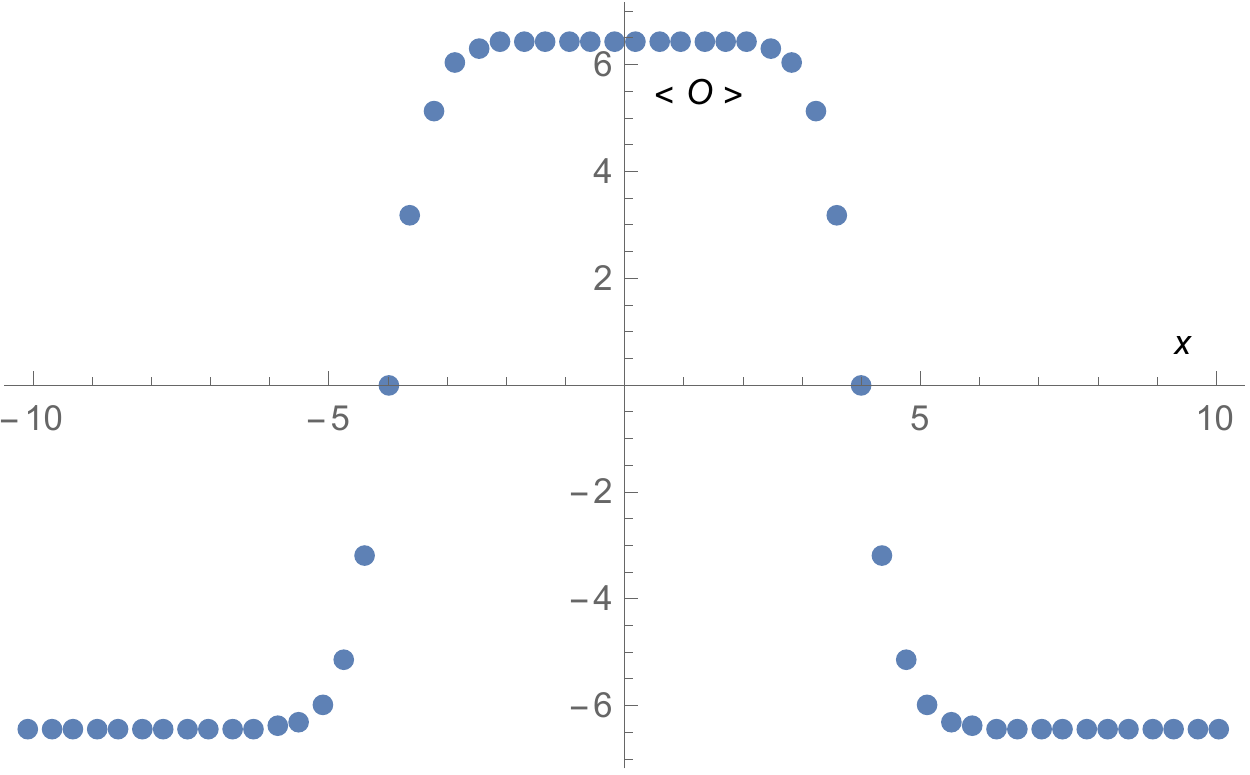}
\includegraphics[scale=0.5]{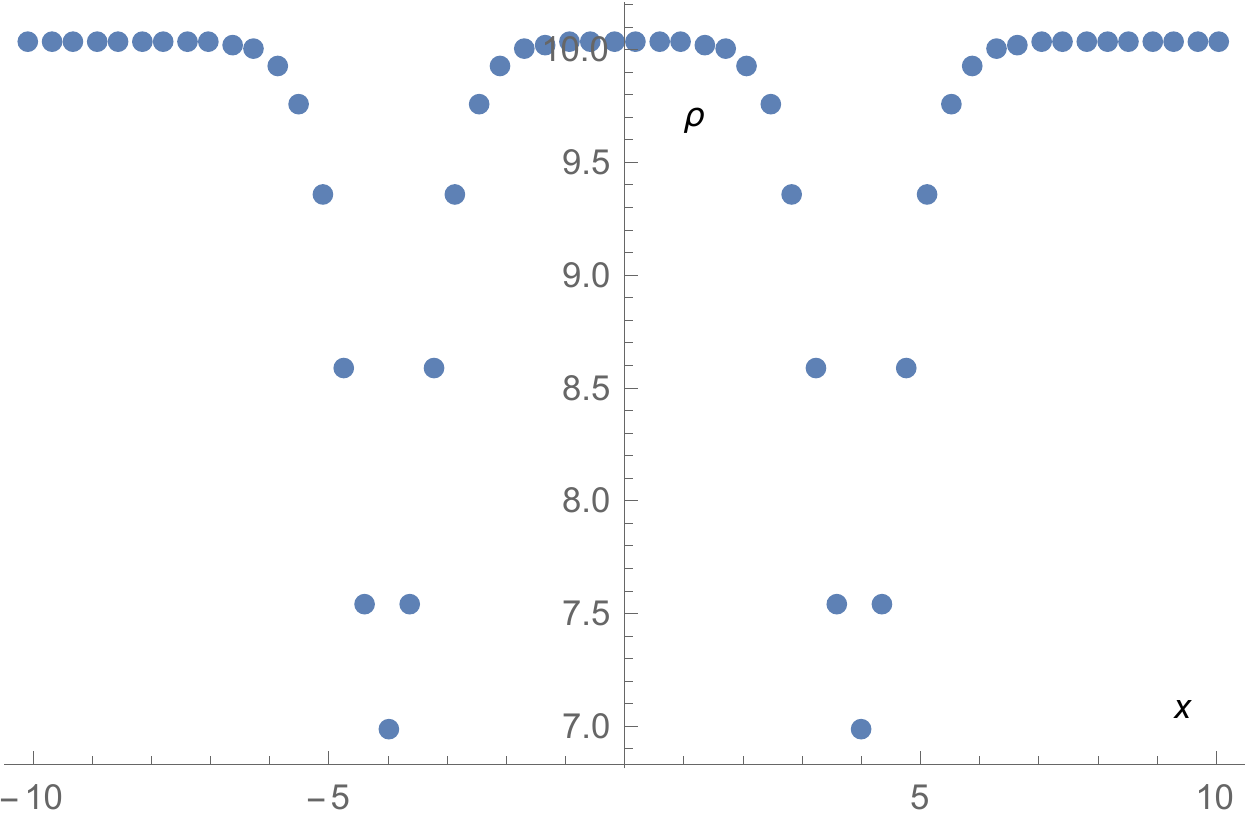}
\end{center}
\caption{Condensate and charge density of the double soliton as functions of $x$ at $\mu=6$.}
\end{figure}

\begin{figure}
\begin{center}
\includegraphics[scale=0.7]{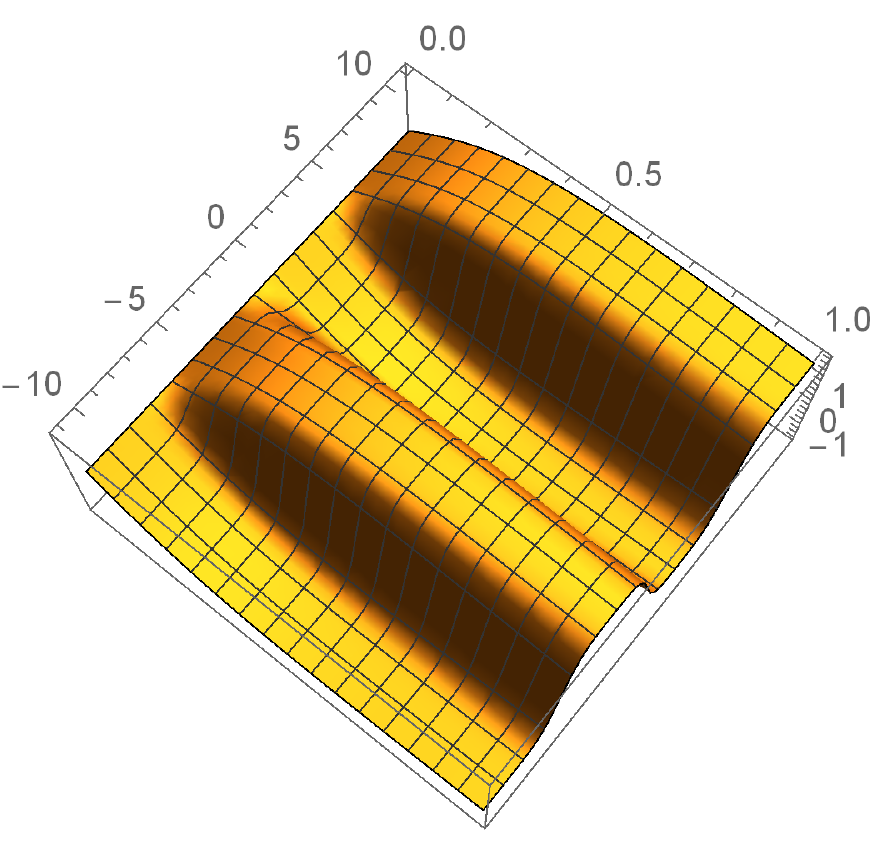}
\includegraphics[scale=0.7]{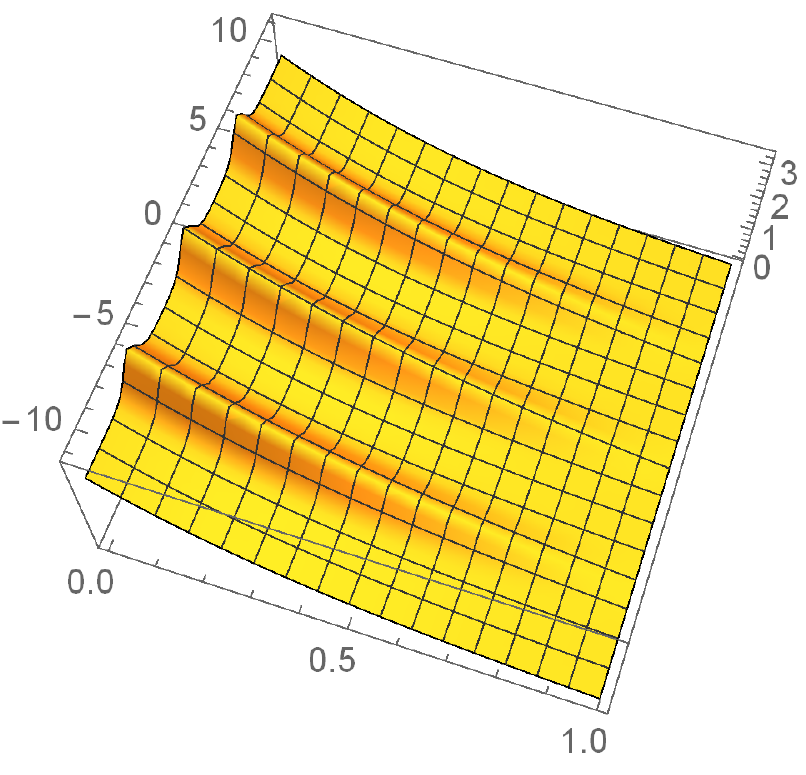}
\end{center}
\caption{The bulk configurations of the triple soliton fields $\phi(x,z),\varphi(x,z)$ from up to down respectively for $\mu=6$.}
\end{figure}
\begin{figure}
\begin{center}
\includegraphics[scale=0.5]{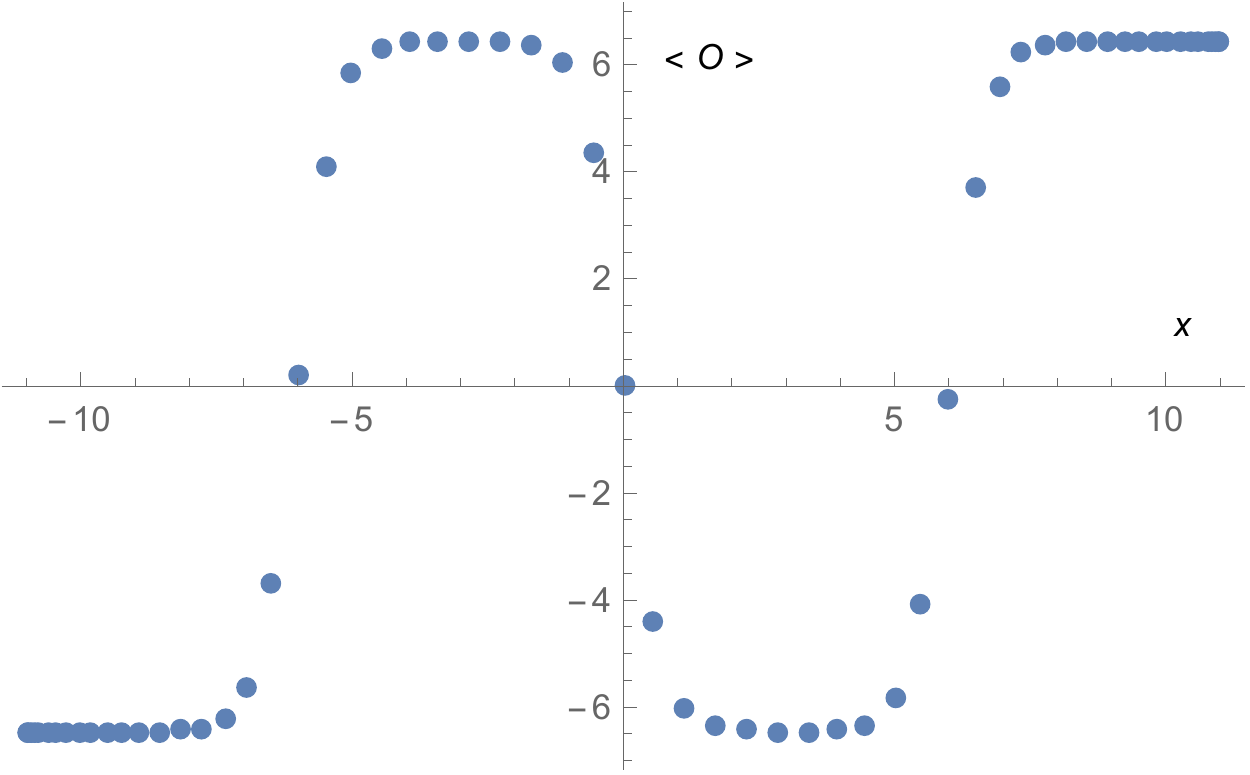}
\includegraphics[scale=0.5]{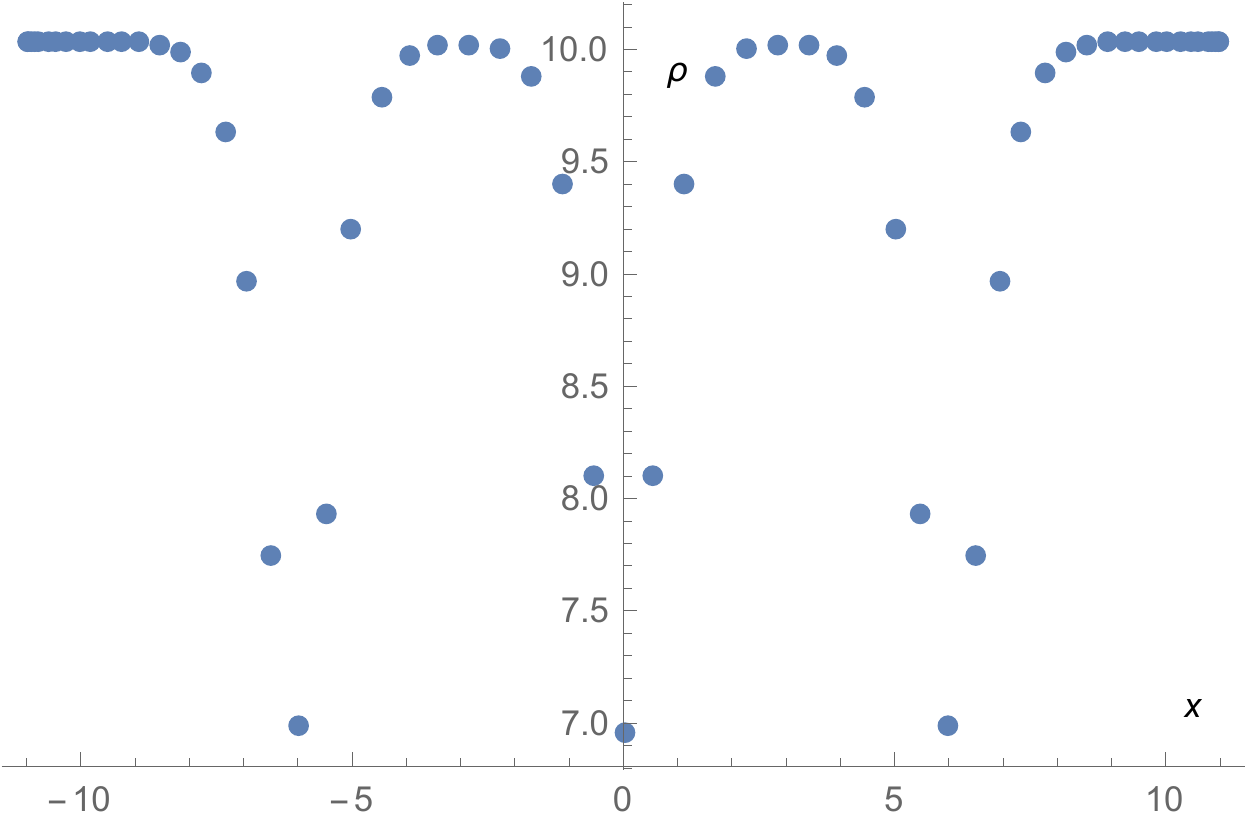}
\end{center}
\caption{Condensate and charge density of the triple soliton as functions of $x$ at $\mu=6$.}
\end{figure}

\bibliographystyle{spphys}
\bibliography{thereference}

\end{document}